\documentclass[
 amsmath,amssymb,
 aps,
 twocolumn
]{revtex4-2}

\usepackage[pdftex]{graphicx}
\usepackage{dcolumn}
\usepackage{caption}
\usepackage{tabularx}
\usepackage{accents}
\usepackage{bm}
\usepackage{braket}
\usepackage{amsmath}
\usepackage{mathrsfs}
\usepackage{booktabs}
\usepackage{array}
\usepackage{hyperref}
\newcommand\Tstrut{\rule{0pt}{2.6ex}}         
\newcommand\Bstrut{\rule[-0.9ex]{0pt}{0pt}}   

\newlength\myindention
\DeclareCaptionFormat{myformat}{\hspace*{\myindention}#1#2#3}
\begin{document}

\title{Variational Monte Carlo Calculations of $n+\text{}^3\text{H}$ Scattering}
\author{Abraham R. Flores}
\author{Kenneth M. Nollett}
\affiliation{San Diego State University, 5500 Campanile Drive, San Diego, CA 92182, USA}
\date{\today}

\begin{abstract}
A paramount goal in nuclear physics is to unify \textit{ab-initio} treatments of bound and unbound states. The position-space quantum Monte Carlo (QMC) methods have a long history of successful bound-state calculations in light systems but have seen minimal implementation in unbound systems. Here we introduce a numerical method to improve the efficiency and accuracy of unbound-state calculations in QMC. As an initial application, we compute scattering observables for the smallest system available to probe three-body forces, the neutron-triton system, using variational Monte Carlo (VMC) wave functions. The method involves inferring long-range amplitudes in the wave function from integrals over the short-range region where all the particles interact. This approach using integral relations is well established in the literature; here, we develop it for the QMC framework. We validate our approach with a consistency check between short-range spectroscopic overlap functions computed from direct evaluation and from the integral relations; scattering amplitudes are long-range asymptotics of those overlaps. Comparison against published benchmark calculations using the same potential demonstrates that when applied to the current VMC wave functions, the integral method produces more accurate scattering observables than direct evaluation from the same variational wave function. However, it still differs noticeably from the exact results. Using additional interactions, we then present phase shifts and mixing parameters for the $n+\text{}^3\text{H}$ system. In particular, we present one of the first applications of the Norfolk family of local coordinate-space chiral potentials in unbound systems of $A > 2$. The Norfolk results accurately describe $s$-wave scattering but predict $p$-wave cross sections too large. Compared with previous QMC scattering calculations, the integral method avoids difficulties associated with the precise computation of energy differences and with convergence outside the interaction region, which is particularly severe in the variational calculation. Application of the integral method here paves the way for its use in Green's function Monte Carlo (GFMC) calculations. In GFMC, the wave functions are more accurate, but the high-precision convergence of their tails is slow, and there are additional difficulties in reading out amplitudes. The integral methods will address both of those remaining problems.
\end{abstract}

\maketitle

\section{\label{sec:intro}Introduction}
 
A unified \textit{ab-initio} description of bound and unbound nuclear systems is a long-standing objective of theoretical nuclear physics \cite{Johnson2020White}. Because unbound wave functions are not easily represented in a basis of manageable size, and desired quantities are not easily reduced to eigenvalue or Rayleigh-Ritz problems, even formulating useful methods beyond the mass-4 system is difficult. As a result, far less progress has been made on \textit{ab initio} calculations of unbound than bound states. As with other computational frameworks, the quantum Monte Carlo (QMC) methods based on position-space sampling \cite{Carlson2015QMC} were developed mainly in a bound-state context, so that the successes of the Green's function Monte Carlo (GFMC) method in nuclear systems so far have been almost entirely in the modeling of bound states \cite{wiringa2000quantum};  the method has received only minimal adaptation for unbound systems, where it has been applied twice to $^5\text{He}$ \cite{nollett2007quantum,lynn2016Chiral}. In this paper, we present and test the application of short-ranged integral relations that reduce the difficulty and increase the accuracy of applying the QMC methods to unbound systems. These integral relations are closely related to the source-term technique of Refs. \cite{pinkston1965Form,kawai1967Amethod}, the ANC calculations of Refs. \cite{mukhamedzhanov1990Micro,timofeyuk1998One,timofeyuk2010Overlap,nollett2011asymptotic,nollett2012ab} and the scattering calculations of Refs. \cite{harris1967expansion,barletta2009integral,kievsky2010variational,Romero2011General,viviani2020n+}, but with adaptation to the context of QMC. 
 
Here we develop and present the improved technique within the context of the four-nucleon ($A = 4$) system. In recent decades, this system has served as a practical testing ground for comparing the accuracy and consistency of computational methods and constraining the three-nucleon interaction. In Ref.~\cite{kamada2001benchmark}, the $^4\text{He}$ ground state was used to benchmark GFMC and six other many-body methods, finding agreement within 1\% on the binding energy when the AV8$^\prime$ potential \cite{pudliner1997quantum} was used. A similar agreement was found in studies that included phenomenological three-nucleon interactions \cite{wiringa2000quantum, nogga2002alpha, lazauskas2004testing, viviani2005calculation}. The four-body bound system is important in determining the three-nucleon interaction through the binding energy of $^4\text{He}$ \cite{Pudliner1995Quantum,Carlson2015QMC,baroni2018local,machleidt2011chiral}.

\textit{Ab initio} calculations of unbound four-nucleon systems have matured significantly in the last decade and a half. In Refs. \cite{viviani2011benchmark, viviani2017benchmark} three \textit{ab-initio} computational methods were benchmarked against scattering observables in the $n/p+\text{}^3\text{H}$ and $n/p+\text{}^3\text{He}$ systems using realistic interactions and finding good agreement between methods. The benchmarked methods were the configuration-space Faddeev-Yakubovsky equations \cite{lazauskas2004testing,lazauskas2005low,lazauskas2009elastic,lazauskas2012application,lazauskas2020description}, the Kohn variational principle in a hyperspherical harmonics basis \cite{kievsky2008high,marcucci2009n,barletta2009integral,kievsky2010variational,viviani2020n+}, and the Alt-Grassberger-Sandhas equations solved in momentum space \cite{deltuva2014calculation,deltuva2015deuteron,deltuva2015four,deltuva2015proton,deltuva2017four,alt1978coulomb,alt1980scattering,deltuva2005momentum,deltuva2005calculation}. The resonating group method (RGM) has also been applied, mostly with effective interactions and restricted bases \cite{hofmann1997microscopic,pfitzinger2001elastic,hofmann2003microscopic,hofmann2008he,Arai2010Micro,Arai2011Tensor,aoyama2012Four}. In addition, the RGM has been merged with the no-core shell model to compute scattering and reactions, but the only application of this approach to $A = 4$ scattering that we are aware of was very early in the development of the method \cite{quaglioni2008ab,navratil2010ab}.

Another important recent development has been the advent of systematically-organized phenomenological potentials based on chiral effective field theory \cite{machleidt2011chiral,piarulli2020local}. For some time, these were all formulated in momentum space in ways that made their use with QMC methods problematic \cite{Gezerlis2014Local,Lynn2012Real}. This has now changed with the development of chiral models suited to position-space calculations \cite{piarulli2015minimally,gezerlis2013quantum,Tews2016Quantum}. The Norfolk family of coordinate-space local chiral interactions \cite{piarulli2016local,baroni2018local}  have been applied successfully to binding energies, electroweak transitions, and other properties of light nuclei, as well as infinite nuclear matter \cite{piarulli2018light,marcucci2020hyperspherical,King2020Weak,King2020Chiral,piarulli2020benchmark,lovato2022benchmark}. However, there are extremely limited results for nuclear scattering with the Norfolk interactions \cite{marcucci2020hyperspherical,viviani2022X17}. The further development of QMC methods for unbound systems will help to evaluate and improve the local chiral potentials.  

The few QMC calculations of nuclear scattering that have been published are all based on representing continuum states with particle-in-a-box states. In Ref. \cite{carlson1984variational}, variational Monte Carlo (VMC) was applied to the $p+\text{}^3\text{H}$ system with an older interaction. In Ref.~\cite{carlson1987microscopic}, a similar calculation was carried out for $n+\text{}^4\text{He}$. In Refs.~\cite{nollett2007quantum,lynn2016Chiral} GFMC was adapted to compute phase shifts for $n+\text{}^4\text{He}$. Below we improve on the formulation of the scattering problem for QMC methods, and we test the new formulation on VMC solutions in the $A = 4$, isospin-1 system. We find that it significantly improves VMC calculations of scattering observables because they come much closer to the results of more-exact methods using the same potentials. However, the main benefit is in developing this formulation to be usable as part of GFMC calculations; GFMC uses VMC wave functions as starting points and improves them by filtering out excited-state contamination via imaginary-time evolution.
 
As in prior VMC and GFMC scattering calculations, we proceed by setting up a particle-in-a-box problem with a logarithmic-derivative boundary condition at the surface of a spherical box.  For single-channel scattering, a positive-energy wave function in the box is the short-range part of a continuum wave function at the same energy; the phase shift can be inferred from the energy and the boundary condition in what we call the ``direct procedure.'' Applying this approach to VMC presents two related points of difficulty that are viewable as a mismatch between the computed energy and the imposed boundary condition. First, VMC, as presently organized, is not as good at optimizing the outer part of the wave function as it is the short-range region where all the nucleons interact. Consequently, the influence of the boundary condition on the computed energy is somewhat indirect and dependent on the variational ansatz, limiting the accuracy of the solution. Second, the accuracy of VMC energies decreases significantly as one proceeds beyond the $0s$ shell, while low-energy scattering is very sensitive to the placement of thresholds so that the comparison of direct VMC scattering calculations with measurements or with more exact calculations is generally rather disappointing. In any case, the computational cost of the direct approach consists entirely of the Monte Carlo integration of energy expectation values, both during minimization with respect to variational parameters and in a final high-statistics calculation of the energy. 

One path around these difficulties is a connection between the well-computed short-range wave functions in VMC and the outer part of the true solution in the form of the integral relations discussed above. In the context of VMC, similar integral relations have been applied to single-nucleon asymptotic normalization coefficients (ANCs), resonance widths, and spectroscopic overlaps \cite{nollett2011asymptotic,nollett2012ab}. These ``integral'' calculations in VMC represent an alternative to the direct approach that combines the short-range accuracy of the VMC ansatz with additional information about the Hamiltonian. Since both approaches can be used to compute spectroscopic overlaps, accurate results from the direct method at short range can validate the integral method and its implementation.

Computing curves of energy-dependent scattering observables in VMC by either the direct or the integral method demands considerable human and machine effort. In principle, each energy and each partial wave requires variational energy minimization of a distinct wave function, followed in the integral method by a high-statistics Monte Carlo integration that includes operations on the wave function by the potential operator. In the $n+\text{}^3\text{H}$ system, there are eight relevant angular-momentum channels at low energy, so the scattering matrix at a given energy requires eight VMC wave functions. Doing this at ten different energies for a single Hamiltonian requires eighty wave functions, all with attendant difficulties in guessing the correct boundary condition for the desired energy and actually computing that energy accurately. Repeating for multiple interactions (e.g., the various Norfolk potentials) only increases the amount of work to be done. However, previous ANC calculations by this method \cite{nollett2011asymptotic,nollett2012ab} point the way to an approximation that is simpler to manage while delivering similar accuracy: for each angular momentum channel, a single wave function optimized to lie in the middle of the interesting energy range can be used in integral relations that assume a range of scattering energies. This procedure works because the integral relations are only sensitive to the short-range wave function, which does not change drastically with scattering energy over a range of a few MeV above threshold (in the absence of resonances); we refer to it as the fixed interior wave approximation (FIW). With the need for only one or two wave functions per channel, this approximation drastically reduces human and computer effort compared with computing separate variational wave functions at every desired energy and extracting scattering information from them.

The remainder of this paper is organized as follows: In Sec.~\ref{sec:wavefunctions} we describe the variational wave functions and their adaptation to scattering states. In Sec.~\ref{sec:scattering-matrices} we connect our wave functions to the scattering formalism and establish notation. In Sec.~\ref{sec:amplitude-methods} we describe how to extract scattering amplitudes and spectroscopic overlaps from the wave function using both the ``direct'' and ``integral'' procedures.  In Sec.~\ref{sec:verfiyIM}, we validate and tune the integral method (including the FIW approximation) by comparing direct and integral overlaps.  In Sec.~\ref{sec:results} we present scattering phase shifts and cross sections for multiple nucleon-nucleon interactions, and compare results against published benchmarks. In Sec.~\ref{sec:conclusion} we summarize our findings and briefly discuss their implications for the future.  In an Appendix \ref{appendix:scattering} we provide additional scattering formalism to avoid ambiguity in the meanings of our results.

\section{Variational wave functions}
\label{sec:wavefunctions}

\subsection{Structure of bound states}
\label{sec:struct-bound-states}

In our calculations we use variational wave functions of the form generally employed for nuclear VMC in bound systems~\cite{carlson1991variational, pudliner1997quantum,lagaris81}, with minor modifications for scattering.  Here we briefly describe the structure of the VMC wave functions, emphasizing aspects that require modification for scattering.

The variational wave function $|\Psi_V\rangle$ is constructed from two- and three-body operator correlations acting on a Jastrow wave function that contains only scalar correlations, assembled to have definite quantum numbers of angular momentum quantum and isospin as well as antisymmetry under particle exchange \cite{Carlson2015QMC}.  The form of the variational ansatz used here is~\cite{wiringa2009}
\begin{equation}
    \label{varwf}
    \Ket{\Psi_V} = \left[\mathcal{S}\prod_{i<j}\left(1+U_{ij}
      +\sum_{k\neq i,j}U_{ijk}\right)\right]\Ket{\Psi_J},
\end{equation}
where the sums and products run over nucleon labels.  The operator correlations contain the same operators on spin, isospin, and coordinates that appear in the six largest terms of the nucleon-nucleon interaction.  They have the forms
\begin{eqnarray}
    \label{paircorr}
    U_{ij} = \sum_{p=2,6}\left[\prod_{k\neq i,j}f^p_{ijk}(r_{ik},r_{jk})\right]u_p(r_{ij})O^p_{ij}, 
\end{eqnarray}
where the indexed operators are $O^{p=1,6}_{ij}=[1,\boldsymbol{\sigma}_i\cdot\boldsymbol{\sigma}_j,S_{ij}]\otimes[1,\boldsymbol{\tau}_i\cdot\boldsymbol{\tau}_j]$.  Here $\boldsymbol{\sigma}_i$, $\boldsymbol{\tau}_i$, and $S_{ij}$ are respectively spin, isospin, and tensor operators, and $r_{ij}$ is the distance between nucleons $i$ and $j$.   The spatial dependence $u_p(r_{ij})$ in each term is computed from two-body Euler-Lagrange equations \cite{wiringa91} that contain both the nucleon-nucleon potential and variational parameters, while the correlations $f^p_{ijk}$ suppress spin-isospin pair correlations when a third particle is nearby.  The three-body correlations $U_{ijk}$ are constructed from operators appearing in three-body terms of the potential.
The operator $\mathcal{S}$ in Eq.~(\ref{varwf}) symmetrizes over orderings of the operators (which do not commute) so that $\ket{\Psi_V}$ inherits the antisymmetry of the Jastrow function.

For a given permutation of particle labels (i.e., before antisymmetrization), each particle in the Jastrow function $|\Psi_J\rangle$ is assigned to the $s$- or the $p$-shell.  The central pair correlations depend on this assignment, so for example $f_{ss}(r_{ij})$ is applied when particles $i$ and $j$ are both in the $s$-shell core described below, while $f_{sp}(r_{ij})$ is applied when one is in the the $s$- and one in the $p$-shell.  Then for a nucleus of $A$ nucleons with a full four-particle $s$-shell and at least one $p$-shell particle (presented to connect our work to notation in the prior literature),
\begin{eqnarray}
    \ket{\Psi_J} &=& \mathcal{A}\left\{ \prod_{i<j<k\leqslant 4}f_{ijk}^{sss}\prod_{t<u\leqslant 4}f^{ss}(r_{tu})\right.\nonumber\\
    \label{jastrow}
    &&\times\prod_{i\leqslant 4}\hspace{1mm}\prod_{5\leqslant j \leqslant A}f^{sp}(r_{ij})\prod_{5\leqslant k < l \leqslant A}f^{pp}(r_{kl})\\
    &&\times\left.\sum_{LS[n]}\beta_{LS[n]}\left|\Phi_A(LS[n]JMTT_3)_P\right\rangle \right\}.\nonumber
\end{eqnarray}
If there is only one $p$-shell particle, then $f_{ij}^{pp}=1$.  The function
\begin{eqnarray}
    \label{phiA}
    &&\left|\Phi_A\left(LS[n]JMTT_3\right)_P\right\rangle\nonumber\\
    &&=\left|\Phi_\alpha(0000)_{1234}\prod_{5\leqslant i\leqslant A}\phi_p^{LS[n]}(r_{\alpha i})\right.\nonumber\\
    &&\times\left[\left[\prod_{5\leqslant j\leqslant A}
        Y_{lm_l}\left(\mathbf{\hat{r}}_{\alpha j}\right)\right]_{LM_L}
      \!\otimes\left[\prod_{5\leqslant k \leqslant A}\chi_k \left(\frac{1}{2}m_i\right) \right]_{SM_S}\right]_{JM}\nonumber\\
    &&\left.\times\left[\prod_{5\leqslant l \leqslant A}\nu_i\left(\frac{1}{2}t_z\right) \right]_{TT_z}\right\rangle,
\end{eqnarray}
is a spin-isospin vector that depends on particle positions, in which the first four nucleons are assigned to the $s$-shell (or ``alpha core'').  This core is constructed as a simple Slater determinant of spins and isospins coupled to definite total angular momentum and isospin quantum numbers denoted by $\Phi_\alpha(J\text{=}0,M\text{=}0,T\text{=}0,T_z\text{=}0)_{1234}$, while the remaining particles are assigned to $p$-shell orbitals.  Spinors $\chi_i$ and $\nu_i$ specify spin and isospin states of the $p$-shell particles,  and spherical harmonics $Y_{lm_l}$ describe their angular motion around the center of mass of the core, from which they are separated by vectors $\mathbf{r}_{\alpha i}$.  These are coupled to specified quantum numbers of total spin $S$, orbital angular momentum $L$, and net angular momentum $J$ (with projection $M$), as well as total isospin $T$ (with projection $T_z$), as indicated by square brackets.  Full specification in general also requires definite permutation symmetry among $p$-shell orbitals, in the form of a Young diagram label $[n]$.

Each $p$-shell orbital $\phi_p^{LS[n]}(r_{\alpha i})$ depends on the magnitude of $\mathbf{r}_{\alpha i}$ (which keeps the wave function translation-invariant), and on the specified quantum numbers.  Since pair correlations appear elsewhere, the $\phi_p^{LS[n]}$ can be thought of as accounting for interactions with the mean field of the nucleus, as well as allowing antisymmetry when $A>4$.  The $f_{sp}$ are constructed explicitly to describe close-in pair correlations, so that $f_{sp}(r\rightarrow\infty) = 1$, and the exponential drop-off of the wave function at large $r_{\alpha i}$ for bound states is built into $\phi_p^{LS[n]}$.  Each $\phi_p^{LS[n]}(r_{\alpha i})$ is accordingly computed from a Woods-Saxon potential well with orbital angular momentum $l=1$ (for a $p$-shell nucleus) in a one-body Schr\"odinger equation.  The geometric parameters of the wells and their separation energies are parameters to be varied in the VMC procedure, with the initial guess for separation energy given by the appropriate breakup threshold for the nuclear system at hand.  Configurations having all of the quantum numbers $L$, $S$, and $[n]$ that can be consistent with the given $J$, $T$, and $T_z$ are in general present, so they are all included with weight amplitudes $\beta_{LS[n]}$ in Eq.~(\ref{jastrow}).  Finally, the operator $\mathcal{A}$ denotes an antisymmetric sum over every permutation $P$ of particle labels.

All of the correlations and orbitals are either explicit functions of variational parameters or else solutions of differential equations whose constants are treated as variational parameters.  The variational wave function is an approximate solution of the Schr\"odinger equation,
\begin{equation}
  \label{eq:schroedinger}
  \hat{H}\Psi = E\Psi
\end{equation}
with Hamiltonian operator $\hat{H}$.
Optimal values of all parameters are found from a Rayleigh-Ritz variational principle by minimizing the energy expectation value
\begin{equation}
\label{energyexpect}
 E_V = \frac{\bra{\Psi_V}\hat{H}\ket{\Psi_V}}{\braket{\Psi_V|\Psi_V}}.
\end{equation}
The integrals in
Eq.~(\ref{energyexpect}) and other matrix elements are computed by
Monte Carlo integration, using $\Psi_V^\dag\Psi_V$ (a function of particle coordinates) as weight function.  We optimize parameters using an implementation of the non-linear optimizing
algorithm COBYLA from the NLopt library~\cite{nlopt}.  After
minimization, the result of VMC is a variational upper bound on the
energy expectation value and an optimized wave function that can
be used as an input to further calculations.

\subsection{Adaptation to scattering states}
\label{sec:vmc-scattering-states}

This general framework is adaptable to unbound states \cite{carlson1984variational,carlson1987microscopic,nollett2007quantum}.  The eigenvalue nature of bound states arises from the condition that their wave functions should be square-integrable.  This is enforced in VMC by exponential decay at large distance in the $f_{ss}$ correlations and in the $p$-shell orbitals $\phi^{LS[n]}_p$.  Scattering formally involves wave functions that extend to infinity and cannot be found by energy minimization.  However, we can adapt the VMC procedure and wave function to scattering by confining the wave function to a spherical box and imposing a boundary condition at its edge.  The boundary condition renders the finite-domain kinetic energy operator Hermitian and the wave function normalizable, so that VMC energy minimization gives access to a unique ground state.  Once this ground state has been found, it may be matched smoothly onto asymptotic scattering solutions outside the box and viewed as the portion of a scattering wave function near the origin.

The most useful boundary condition for single-channel scattering is a specified logarithmic derivative $\zeta_c$ of the wave function at the box surface, defined by
\begin{equation}
  \label{boundarycond}
  \mathbf{\hat{n}}_c\cdot\nabla_{\mathbf{r}_c}(r_c \Psi) = \zeta_cr_c\Psi,
\end{equation}
where the subscript $c$ denotes a scattering channel specified by a division of nucleons into two nuclei and by values of $J,M_J,L$, and $S$.   The gradient is evaluated in coordinates defined by the vector $\mathbf{r}_c$  separating the centers of mass of the scattering nuclei, and $\mathbf{\hat{n}_c}$ is an outward normal unit vector at the box surface, defined by $r_c=R_0$ for a box of radius $R_0$.  For fixed $R_0$, different choices of $\zeta_c$ give boxes with different ground state energies, corresponding to scattering at different energies.

The wave function ansatz of Eqs.~(\ref{varwf})--(\ref{phiA}) is easily adapted to describe scattering of a single nucleon by an $s$-shell nucleus as in our $n+\,^3\mathrm{H}$ case.  The $s$-shell nucleus is well-described by a spin-isospin Slater determinant and correlations between its nucleons, just like the $s$-shell portion of Eqs.~(\ref{varwf}) and (\ref{jastrow}) \cite{carlson1983three-nucleon}.  The scattered nucleon can be incorporated into the wave function just like the $p$-shell particles in Eq.~(\ref{phiA}), using the same routines in the VMC code but narrowing to the case of only one ``$p$-shell'' particle, which could have any value of $L=l$ for its orbital motion around the nuclear center of mass, not just $l=1$.  Then the coordinate $\mathbf{r}_{\alpha j}$ in Eq.~(\ref{phiA}) is identical to the channel separation $\mathbf{r}_c$ for scattering.  For scattering from a triton, the  core contains only three particles, coupled to $J=1/2$, $T=1/2$, $T_z=-1/2$ to form $\Phi_t(\frac{1}{2}\frac{1}{2}\frac{1}{2}(-\frac{1}{2}))_{123}$; the $ss$ and $sss$ correlations should in principle be optimized for the triton ground state so that configurations at the box surface truly match onto a triton cluster outside the box.  To produce a specific scattering channel with good quantum numbers, the angular momentum coupling in Eq.~(\ref{phiA}) has to be adapted to couple the scattered nucleon to the $J=1/2$ core.  Fermionic exchange of the scattered nucleon with a nucleon from the core enters through the antisymmetrization in Eq.~(\ref{jastrow}).  A similar but more elaborate approach has been used for scattering of composite nuclei and for highly-clusterized nuclei, when one of the clusters is an alpha particle \cite{nollett2001li6,nollett2001li7be7,wiringa2002weak}.

Since the $sp$ pair correlations are constrained to go over to the identity operator at large separation, placement of $R_0$ at large enough radius turns Eq.~(\ref{boundarycond}) into a boundary condition on computation of $\phi_p^{LS[n]}$ from its Woods-Saxon well, so that in principle
\begin{equation}
  \label{phiboundarycond}
 \zeta_cr_c\phi_c(R_0) = \left.\frac{d(r_c\phi_c)}{dr_c}\right|_{r_c=R_0}.
\end{equation}
However, in optimizing the wave function it is often possible to lower $E_V$ by altering the cutoff parameter that enforces $f_{sp}\longrightarrow 1$ at large separations, as assumed in using Eq.~(\ref{phiboundarycond}) as a boundary condition on the whole wave function.  When the cutoff radius becomes too large, the $f_{sp}$ correlations gain a slope at $R_0$ and alter $\zeta_c$. To solve this problem we apply the condition in Eq.~(\ref{phiboundarycond}) to the product $[f_{sp}(r_c)]^{n_t}\phi_c(r_c)$ instead of just $\phi_c(r_c)$ by itself,  where $n_t$ is the number of nucleons in the scattering nucleus ($n_t=3$ for $n+\,^3\mathrm{H}$).  This works because high-probability configurations all have the distance from a neutron at the box boundary to any nucleon inside the triton (mean radius 1.7 fm, smaller than our 9-fm box) close to $r_c$.  We found that the modified Eq.~(\ref{phiboundarycond}) enforces the desired boundary condition in Eq.~(\ref{boundarycond}) on the wave function with good precision.  It also prevents the optimizer from pushing the $f_{sp}$ cutoff to larger radius, removing a significant source of difficulty in the variational search.

For coupled-channels problems there is a $\zeta_c$ for each channel, and also a $\beta_c$ amplitude (corresponding to the $\beta_{LS[n]}$ of a bound state).  The values of $\zeta_c$ determine the energy of the lowest state in the box, which we find by the usual energy minimization.

\section{Wave functions outside the box}
\label{sec:scattering-matrices}

At the boundary, the wave function projection into a specific channel matches onto a solution of the Coulomb wave equation.  This is a radial Schr\"odinger equation with a Coulomb potential and positive energy, and in dimensionless form it is
\begin{equation}
     \label{coulombscrhodinger}
    -\frac{d^2u_l}{d\rho^2} +\left(\frac{l(l+1)}{\rho^2}+\frac{2\eta}{\rho}\right)u_l = u_l.
\end{equation}
Here $\rho = k_cr_c$, while the Sommerfeld parameter is given in terms of the charges $Z_1e$ and $Z_2e$ and the reduced mass $\mu$ by $\eta = Z_1Z_2e^2\mu/(\hbar^2 k_c)$. The channel wave number is given in terms of the channel energy $E_c$ by $k_c^2 = 2\mu E_c/\hbar^2$.   Real-valued independent solutions of Eq.~(\ref{coulombscrhodinger}) are the usual regular $F_l(\eta, \rho)$ and irregular $G_l(\eta, \rho)$ Coulomb functions \cite{NIST:DLMF}.  (For neutron scattering $\eta=0$, and these are equivalent to spherical Bessel functions; we will retain the more-general notation of the Coulomb functions.)  Outside the interaction region, the wave function may be written in terms of products of these functions with products of wave functions for the individual colliding nuclei, coupled to specified angular momentum quantum numbers.  These products form the regular and irregular channel-cluster functions
\begin{equation}
    \label{Fcompact}
    \mathcal{F}_c = \Psi_{1\otimes2}^c\frac{F_{l_c}(\eta_c,k_cr_c)}{k_cr_c}
\end{equation}
and
\begin{equation}
    \label{Gcompact}
    \mathcal{G}_c = \Psi_{1\otimes2}^c\frac{G_{l_c}(\eta_c,k_cr_c)}{k_cr_c}.
\end{equation}
We define the channel product function of cluster wave functions $\psi_{1c}$ and $\psi_{2c}$ with specified angular momentum in channel $c$ as
\begin{equation}
    \label{12wf}
    \Psi^c_{1\otimes2} = \mathcal{A}_c\left[\psi_{1c}^{J_{1c}}\otimes\left[\psi_{2c}^{J_{2c}}\otimes Y_{l_c}(\hat{r}_c)\right]_{j_c}\right]_J.
\end{equation}
The operator $\mathcal{A}_c$ antisymmetrizes the function with respect to partitions of nucleons into the two clusters of channel $c$, which have wave functions $\psi_{1c}^{J_{1c}}$ and $\psi_{2c}^{J_{2c}}$.  The angular momentum coupling in Eq.~(\ref{12wf}) organizes the spin and orbital angular momenta ($J_{2c}$ and $l_c$) of cluster 2 to total $j_c$ (corresponding to ``$jj$ coupling'' when cluster 2 is a single nucleon as in our case.)  This is then coupled to the angular momentum $J_{1c}$ of cluster 1, which in our case is the triton.  We follow this coupling scheme in our calculations because of its long use in the computation of spectroscopic factors and overlaps in QMC \cite{forest1996overlaps,lapikas1999escattering,wuosmaa2005li9,wuoasmaa2005he7,brida2011quantum}; afterward we tranform results to the coupling scheme customary for $n+\,\mathrm{H}$ scattering.

Of the standard scattering-matrix formulations, the one written entirely in terms of standing waves is the $K$-matrix, where the long-range part of the wave function is written as
\begin{equation}
    \label{Kaysmptote}
    \Psi(\text{all } r_c \to \infty)
    = \sum_c \left(A_c\mathcal{F}_c + B_c\mathcal{G}_c\right).
\end{equation}
The amplitudes $A_c$ and $B_c$ determine all scattering observables, so the task of theoretical calculations is to find relations among them across all channels.   In  single-channel scattering there is only one relation between amplitudes, expressible in terms of the phase shift $\tan\delta = B/A$.  When there are coupled channels, the $K$-matrix gives the set of $B_c$ coefficients in terms of the $A_c$ coefficients.

While our many-body computation is entirely in terms of the $A_c$ and $B_c$ parameters of the $K$-matrix formalsim, several aspects of scattering are more naturally expressed in the $T$- or $S$-matrix formalism.  In  Appendix \ref{appendix:scattering} we describe the relationships between the scattering matrices, their connections to the computed amplitudes, and the conventions used for phase shifts and mixing parameters in $n+\,^3\mathrm{H}$ scattering.

\section{Determination of amplitudes}
\label{sec:amplitude-methods}
\subsection{The Direct Method}
\label{sec:direct-method}

Applying  Eqs.(\ref{Kaysmptote}), (\ref{boundarycond}) and (\ref{blatt}) at the boundary $R_0$ (and omitting the channel label $c$) one finds $B/A=\tan \delta$ for single-channel scattering.  It then follows from Eqs.~(\ref{boundarycond}) and (\ref{Kaysmptote}) that $B/A$ is equal to
\begin{eqnarray}
\label{tandSC}
  &&\tan\delta = \left .\frac{k\frac{\partial}{\partial \rho}F_{l}(\eta,\rho)-\zeta F_{l}(\eta,\rho)}{\zeta G_{l}(\eta,\rho)-k\frac{\partial}{\partial \rho}G_{l}(\eta,\rho)}\right|_{\rho = kR_0}.
\end{eqnarray}
Here we have the phase shift as an explicit function of $\zeta$ and of the channel energy (which enters through $\eta$ and $k$), with no need to compute $A$ or $B$.  Choosing a value of $\zeta$, we compute the corresponding energy, subtract the threshold energy to obtain the channel energy, and compute the phase shift from Eq.~(\ref{tandSC}).  Repetition at several $\zeta$ values maps out $\delta$ as a function of energy.  When there are coupled channels, the linear relation among amplitudes is more complicated than a simple ratio, and one of Eqs.~(\ref{kmat}), (\ref{tmat}), or (\ref{smat}) has to be inverted.  In computing a wave function for that case,  the $\zeta_c$ values for all channels are inputs.  Inversion of Eq.~(\ref{kmat}) to obtain $\hat{K}$ requires two linearly independent solutions constructed (using different sets of $\zeta_c$) to have the same or nearly the same energy.

The amplitudes $A_c$ and $B_c$ are effectively values of spectroscopic overlap functions at the box surface, defined by projection of a state onto the cluster-product function of Eq.~(\ref{12wf}).  A spectroscopic overlap onto a scattering channel is defined as
\begin{equation}
  \label{directR}
  R_c(r) = \frac{1}{\mathcal{N}}
  \bra{\Psi^c_{1\otimes 2}}\frac{\delta(r-r_{c})}{r^2_{c}}\ket{\Psi_V},
\end{equation}
where 
\begin{equation}
    \label{normR}
    \mathcal{N} = \sqrt{\braket{\psi_{1c}^{J_{1c}}|\psi_{1c}^{J_{1c}}}\braket{\Psi_V|\Psi_V}}
\end{equation}
normalizes the wave functions (cluster 2 being only a neutron spinor in our $n+^3$H case).  In principle the cluster product function is antisymmetrized with respect to nucleon exchange between clusters, as in Eq.~(\ref{12wf}).  However, the explicit antisymmetry of $|\Psi_V\rangle$ has the result that an antisymmetrized $\ket{\Psi^c_{1\otimes 2}}$ gives the same $R_c(r)$ as a single permutation, multiplied by the square root of the number of possible exchanges ($A$ for nucleon scattering); we therefore use a single permutation and replace the $\mathcal{A}_c$ operator by $\sqrt{A}$.  Finally, all of our wave functions are explicitly translation-invariant, so no correction factor is needed to account for use of fixed-center basis states.

For direct determination of the $A_c$ and $B_c$ we separate the overlap at the box surface into $F_l$ and $G_l$ terms using its value and derivatative.  As we show below, overlap functions near the center of the box also provide important tests of the integral method, particularly since the routines for one-nucleon removal overlaps have a long history of previous use \cite{lapikas1999escattering,brida2011quantum}.
\subsection{The Integral Method}
\label{sec:integral-method}

Alternatively, the surface amplitudes may be computed from integrals over the part of the box interior where all of the nucleons interact.  We start in the $K$-matrix formalism, defined by the standing wave in Eq.~(\ref{Kaysmptote}).  Application of Green's theorem over a sphere of fixed cluster separation $r_c$, together with the Wronskian relation for the Coulomb functions at the sphere's surface, gives
\begin{equation}
    \label{wronk}
    \frac{2\mu k_c}{\hbar^2}\left(\Bra{\mathcal{F}_c}\hat{H}-E\ket{\mathcal{G}_c}
      - \Bra{\mathcal{G}_c}\hat{H}-E\ket{\mathcal{F}_c} \right)= 1,
\end{equation}
where $H$ is the full many-body Hamiltonian and $E$ is the total energy.  The Dirac bracket here, as elsewhere, denotes full contraction of the spin-isospin vector at every point and integration over all independent nucleon coordinates in the center-of-mass frame.  This expression is nonzero because the Laplacian operator inside $\hat{H}$ is not Hermitian on a finite region (without further specification of a boundary condition) \cite{kievsky2010variational,Romero2011General}.

If the sphere of constant $r_c$ is located where the limit in Eq.~(\ref{Kaysmptote}) holds, we obtain from Eq.~(\ref{wronk})
\begin{equation}
   \label{fullA}
   A_c = \frac{2\mu k_c}{\hbar^2}\left(\Bra{\Psi}\hat{H}-E\ket{\mathcal{G}_c} - \Bra{\mathcal{G}_c}\hat{H}-E\ket{\Psi} \right) 
\end{equation}
and
\begin{equation}
  \label{fullB}
  B_c = \frac{2\mu k_c}{\hbar^2}\left(\Bra{\mathcal{F}_c}\hat{H}-E\ket{\Psi}
  - \Bra{\Psi}\hat{H}-E\ket{\mathcal{F}_c} \right). 
\end{equation}
Applying the condition that $(H-E)\ket{\Psi}=0$ to the surface amplitudes of Eqs. (\ref{fullA}) and (\ref{fullB}) gives
\begin{equation}
\label{Aexact}
   A_c = \frac{2\mu k_c}{\hbar^2}\Bra{\Psi}\hat{H}-E\ket{\mathcal{G}_c}
\end{equation}
and
\begin{equation}
\label{Bexact}
   B_c = -\frac{2\mu k_c}{\hbar^2}\Bra{\Psi}\hat{H}-E\ket{\mathcal{F}_c}.
\end{equation}
The main difficulty in evaluating these expressions lies in Eq.~(\ref{Aexact}), where the divergence in $G_l$ at $r_c=0$ gives rise to a delta-function in the $\nabla^2G_l$ term of the integrand.  We resolve this difficulty using the regularization strategy described in Refs.~\cite{kievsky2010variational,Romero2011General,viviani2017benchmark,viviani2020n+}. The essential property of $G_l$ for application of Eqs.~(\ref{wronk})-(\ref{Bexact}) is that it satisfies a Wronskian relation with $F_l$ outside the interaction region.  Any function satisfying that relation at $r_c=R_0$ would work just as well, so we replace $\mathcal{G}_l$ with a regularized function $\widetilde{\mathcal{G}}_l$ that has the properties
\begin{eqnarray}
    &&\widetilde{\mathcal{G}}_c = f^c_{\text{reg}}\mathcal{G}_c\label{freg}\\
    &&\widetilde{\mathcal{G}}_c(r_c\to 0) = 0\label{freg0}\\
    &&\widetilde{\mathcal{G}}_c(r_c\to R_0) = \mathcal{G}_c \label{fregR0}.
\end{eqnarray}
One possible choice of the regularizer that satisfies these properties for all partial waves and eliminates the delta function is
\begin{equation}
\label{reg}
f^c_{\text{reg}}(\gamma, r) = (1 - e^{-\gamma r})^{2l+1},
\end{equation}
where $\gamma$ is a parameter to be fixed. We tried a few different regularizing functions, but this one produced the best evidence of producing correct results when $\gamma$ is chosen within a favorable range (as shown below). This specific regularizer has been used in hyperspherical harmonics calculations in recent years \cite{kievsky2010variational,Romero2011General,viviani2020n+,marcucci2020hyperspherical}. Here we explore its application in the VMC context.

Starting from Eq.\ (\ref{wronk}) and replacing $\mathcal{G}_l$ with $\widetilde{\hat{G}}_l$ we find similar results as before, with initially
\begin{equation}
\label{regW}
    \frac{2\mu k_c}{\hbar^2}(\Bra{\mathcal{F}_c}\hat{H}-E\ket{\widetilde{\mathcal{G}}_c} - \bra{\widetilde{\mathcal{G}}_c}\hat{H}-E\ket{\mathcal{F}_c} )= 1.
\end{equation}\hspace{1mm}\\
Picking up from there, Eq.~(\ref{Bexact}) for $B_c$ remains the same. However, $A_c$ of Eq.(\ref{Aexact}) becomes 
\begin{equation}
\label{AregExact}
    A_c = \frac{2\mu k_c}{\hbar^2}\Bra{\Psi}\hat{H}-E\ket{\widetilde{\mathcal{G}}_c}.
\end{equation}

Further simplification follows from a separation of the Hamiltonian into three parts, $\hat{H}$=$\hat{H}_\mathrm{rel}+\hat{H}_1+\hat{H}_2$, suggested by partitioning the nucleons of $\tilde{\mathcal{G}}_l$ into clusters $\psi_{c1}$ and $\psi_{c2}$.  Parts $\hat{H}_1$ and $\hat{H}_2$ contain only the relative coordinates and spinors inside $\psi_{c1}$ and $\psi_{c2}$, respectively.  The third part $\hat{H}_\text{rel}$ contains the kinetic energy of cluster relative motion and all terms of the nucleon-nucleon potential acting between nucleons that are not in the same cluster \cite{nollett2012ab}.  We denote this sum of different-cluster potential terms in channel $c$ as $\hat{V}_\text{rel}^c$ (which depends on how nucleons are partitioned).  We similarly divide the energy into $E=E_c+E_1+E_2$, where $E_c$ is energy relative to the threshold of channel $c$, while $(\hat{H}_1-E_1)\psi_{c1}=0$ and $(\hat{H}_2-E_2)\psi_{c2}=0$.  For exact solutions, the $\hat{H}_1$ and $\hat{H}_2$ terms then cancel out of Eqs.~(\ref{Bexact}) and (\ref{AregExact}).  In addition, the point-coulomb potential $V_\mathcal{C}^c\equiv Z_1Z_2/r_c$ can be added to and subtracted from $H$ to take advantage of the appearance of that term in Eq.~(\ref{coulombscrhodinger}) defining the Coulomb functions.
  
\begin{widetext}

After that work and after applying the Laplacian operator in $H_\mathrm{rel}$ to the function $f^c_\text{reg}(r)$ inside $\tilde{G}_l=f^c_\text{reg}G_l$, the integrals of Eqs.~(\ref{Bexact}) and (\ref{AregExact}) become
\begin{equation}
\label{Bint}
B_c
=-\frac{2\mu}{\hbar^2}
\displaystyle\int_{0}^{\infty}\Psi^\dagger(\hat{V}^c_{\text{rel}}-V^c_{\mathcal{C}})
\mathcal{F}_c\,d^{3A}R
\end{equation}
and
\begin{equation}
\label{Aint}
A_c=\displaystyle\int_{0}^{\infty}
\Psi^\dagger\left\{\frac{2\mu}{\hbar^2}(\hat{V}^c_{\text{rel}}-V^c_{\mathcal{C}})\widetilde{\mathcal{G}}_c
-2\frac{d}{dr_c}[f^c_{\text{reg}}(r_c)]\frac{\partial}{\partial\rho_c}[G_{l_c}(\eta_c,k_cr_c)]\frac{\Psi^c_{1\otimes 2}}{r_c}
-\frac{d^2}{dr_c^2}[f^c_{\text{reg}}(r_c)]\mathcal{G}_c\right\}\,d^{3A}R.
\end{equation}
\end{widetext}
Integration in these expression takes place over the coordinates of all nucleons, $\mathbf{R}_i$ with $i=1,...,A$, in the center-of-mass frame.  Once those coordinates are given, the value of $\mathbf{r}_c$ depends on the partition of nucleons into $\psi_{c1}$ and $\psi_{c2}$, so that the cluster antisymmetrizer $\mathcal{A}_c$ in $\Psi^\dag$ must apply to everything to its right in these integrands.  As with the overlap calculation (see discussion below Eq.~(\ref{directR})), we avoid explicit antisymmetrization by choosing only one partition for each Monte Carlo sample and multiplying the final result by the square root of the number of permutations. The integral relations for computing the surface amplitudes $A_c$ and $B_c$ are then readily applied using wave functions $\Psi$, $\psi_{1c}$, and $\psi_{2c}$ computed from VMC.  We evaluate integrals over the entire box interior at $r_c < 9$ fm using Monte Carlo importance sampling with weight function $\Psi^\dag\Psi$; the short range of $\hat{V}_\mathrm{rel}^c-V_\mathcal{C}^c$ ensures that the integrand is zero in the outer parts of the box.

It is implicit in deriving the integral relations that $\Psi$, $\psi_{1c},$ and $\psi_{2c}$ are exact eigenfunctions of their respective Hamiltonians.  Our $\Psi_V$ and $\psi_{1c}$, on the other hand, are variational approximations to those eigenfunctions.  Previous experience in using integral relations with VMC wave functions \cite{nollett2011asymptotic,nollett2012ab} supports their use despite that apparent shortcoming, because it produces results that compare well with experiment.  Much of the utility of the method in fact arises from the circumstance that the interior part of $\Psi_V$, where the VMC ansatz is most successful, is the only part that contributes to the integrals.  We attempted to estimate the size of the error due to  deviations from $(H_1-E_1)\psi_{1c} =0$, but we found that the Monte Carlo sampling variances on those deviation terms swamped their actual size.  Further progress on that question will presumably require modification of the sampling; past experience has essentially always been that proposed modifications are even worse than the standard sampling (e.g.~Ref.~\cite{nollett2001li7be7}), so we did not pursue the question further.
 
As mentioned in Sec.~\ref{sec:intro}, Eqs.~(\ref{Aint}) and (\ref{Bint}) may be viewed as $r\rightarrow \infty$ properties of the overlap of the wave function onto channel $c$.  A generalized version of Eq.~(\ref{Kaysmptote}) applicable at all radii is
\begin{eqnarray}
\label{intR}
&&R_c(r) = \nonumber\\
&&
\frac{1}{\mathcal{N}r}\left\{\bar{A}_c(r)F_{l_c}(\eta_c,k_cr)
+\bar{B}_c(r)G_{l_c}(\eta_c,k_cr)\right\},
\end{eqnarray}
with
\begin{equation}
\label{Aofr}
\bar{A}_c(r)
\equiv\frac{2\mu}{\hbar^2}
\displaystyle\int_{r_c<r}\Psi^\dagger(\hat{V}^c_{\text{rel}}-V^c_{\mathcal{C}})
\mathcal{G}_c\,d^{3A}R
\end{equation}
\begin{equation}
\label{Bofr}
\bar{B}_c(r)
\equiv-\frac{2\mu}{\hbar^2}
\displaystyle\int_{r_c<r}\Psi^\dagger(\hat{V}^c_{\text{rel}}-V^c_{\mathcal{C}})
\mathcal{F}_c\,d^{3A}R,
\end{equation}
so that $\bar{A}_c(r\rightarrow\infty)=A_c$ and $\bar{B}_c(r\rightarrow\infty)=B_c$ (cf.\ Eq.~(34) of Ref.~\cite{nollett2012ab}).  Since Eq.~(\ref{Aofr}) contains the same singularity that motivates the regularizer, the same problem has to be avoided here.  We do that by computing the integral over all space using the regularizer (Eq.~(\ref{Aint})) and then subtracting the portion of the unregularized integral located at larger radius,
\begin{equation}
  \label{eq:practical-a-bar}
\bar{A}_c(r)
 = A_c - \frac{2\mu}{\hbar^2}
\displaystyle\int_{r_c>r}\Psi^\dagger(\hat{V}^c_{\text{rel}}-V^c_{\mathcal{C}})
\mathcal{G}_c\,d^{3A}R.
\end{equation}

Eqs.~(\ref{Aofr}) and (\ref{Bofr}) or their equivalents have been used in the literature to compute cluster overlaps for Hartree-Fock as well as VMC wave functions \cite{philpott1968Some,kawai1967Amethod,nollett2012ab}.  In Ref.~\cite{nollett2012ab} it was found that for VMC wave functions Eq.~(\ref{intR}) produces overlap functions that agree well with Eq.~(\ref{directR}) at $r_c \lesssim 5$ fm, but that diverge from the direct calculation  at large radius by going over to the correct asymptotic shapes for the specified $E_c$.  This result can be interpreted as providing an extension (in a given channel) of the accurate short-range part of the VMC ansatz into parts of the wave function that are more difficult to compute accurately with VMC.  Recall that the integrals in Eqs.~(\ref{Aexact}), (\ref{Bexact}), (\ref{Aofr}), and $(\ref{Bofr})$ are short-ranged because the nucleon-nucleon pair and triplet interactions inside $\hat{V}_\text{rel}^c$ are short-ranged, and (for charged-particle cases) $V_\mathcal{C}^c$ removes the monopole Coulomb interaction at large radius.  For the present calculations, agreement between Eqs.~(\ref{directR}) and (\ref{intR}) at small $r$ is an important tool for code validation and interpretation of results.  

\subsection{Implementation}
\label{sec:implementation}

Based on the above discussion, the procedure to compute scattering observables from VMC via the integral method begins by computing wave functions that minimize Eq.~(\ref{energyexpect}) separately for a scattering state and for the individual colliding nuclei.  This establishes the channel energy $E_c$ corresponding to the imposed boundary conditions, and that $E_c$ is used with the corresponding wave functions to evaluate Eqs.~(\ref{Bint}) and (\ref{Aint}).  Scattering observables are computed from the resulting amplitudes.  To obtain results at multiple energies, one repeats this procedure for many different boundary conditions that yield different $E_c$.

The dependence on $E_c$ in $\mathcal{F}_c$ and $\mathcal{G}_c$ leads to an alternative approximation procedure that avoids repeated energy minimizations.  This approach builds on previous experience imposing experimental separation energies (or resonance energies) on the integral relations in Refs.~\cite{nollett2011asymptotic,nollett2012ab} even when they differed from computed energies.  If there are no sharp resonances, the small-$r_c$ part of a scattering wave function changes very little over an energy range of several MeV above threshold; this is a consequence of $E_c$ being small relative to the potential strength and (at least in some cases) of antisymmetry constraints on the wave function.  Then we expect that nearly all evolution of $A_c$ and $B_c$ amplitudes with energy comes from the dependence of $F_l$ and $G_l$ inside Eqs.~(\ref{Bint}) and (\ref{Aint}) on the channel energy $E_c$.

Below, in addition to finding separate $\Psi$ at every $E_c$ as described above, we also carry out calculations with the following modified procedure:  in each channel, we optimize a single variational wave function with a boundary condition that gives $E_c$ close to \mbox{3.0 MeV}.  We then assume this wave function to be approximately valid throughout the low-energy region, and we apply the integral relations using $F_l$ and $G_l$ computed from many values of  $E_c$ ranging from zero to several MeV above threshold.  This yields scattering observables as functions of the assumed energy, which we compare with benchmarked exact results using the same nucleon-nucleon potential.  We refer to this approach below as the ``fixed interior wave approximation'' for reasons that are apparent on examination of the corresponding overlap functions.
\section{Integral method verification and regularizer choice}
\label{sec:verfiyIM}

Comparing overlap functions computed using Eq.(\ref{directR}) and  Eq.(\ref{intR}) tests our implementation of the integral relations and identifies useful values of the regularization parameter $\gamma$.  We carry out this test by examining states in the $n+\,^3\mathrm{H}$ system computed from the AV18 potential alone, with the boundary conditions $\zeta_c$ chosen to give optimized variational energies $E_V$ (c.m.) in the neighborhood of 3.0 MeV.  At this energy there is a broad resonant structure in the $p$-wave cross section.  

The direct overlap calculation of Eq.~(\ref{directR}) is carried out in the VMC code as a single Monte Carlo integration over all particle coordinates, in which the radial overlap integral at each $r_{tn}$ is computed from the accumulated samples in a spherical shell of finite (0.1 fm) thickness \cite{brida2011quantum}; this accumulation into shells is an implementation of the delta function in Eq.~(\ref{directR}).  In the figures, each $r_{tn}$ is identified as the midpoint of its shell.  The integral-relation overlap of Eq.~(\ref{intR}) is evaluated in a similar procedure carried out at the same time.  The value of $A_c$ is obtained from the regularized integral over all space in Eq.~(\ref{Aint}), and contributions to the integrals in Eqs.~(\ref{eq:practical-a-bar}) and (\ref{Bofr}) at any $r_{tn}$ consist of all Monte Carlo samples at larger radius. The optimized energy $E_V$ is used initially to specify $k_c$ and $\eta_c$ in these integral calculations.

\begin{figure}
\includegraphics[width=8.6cm]{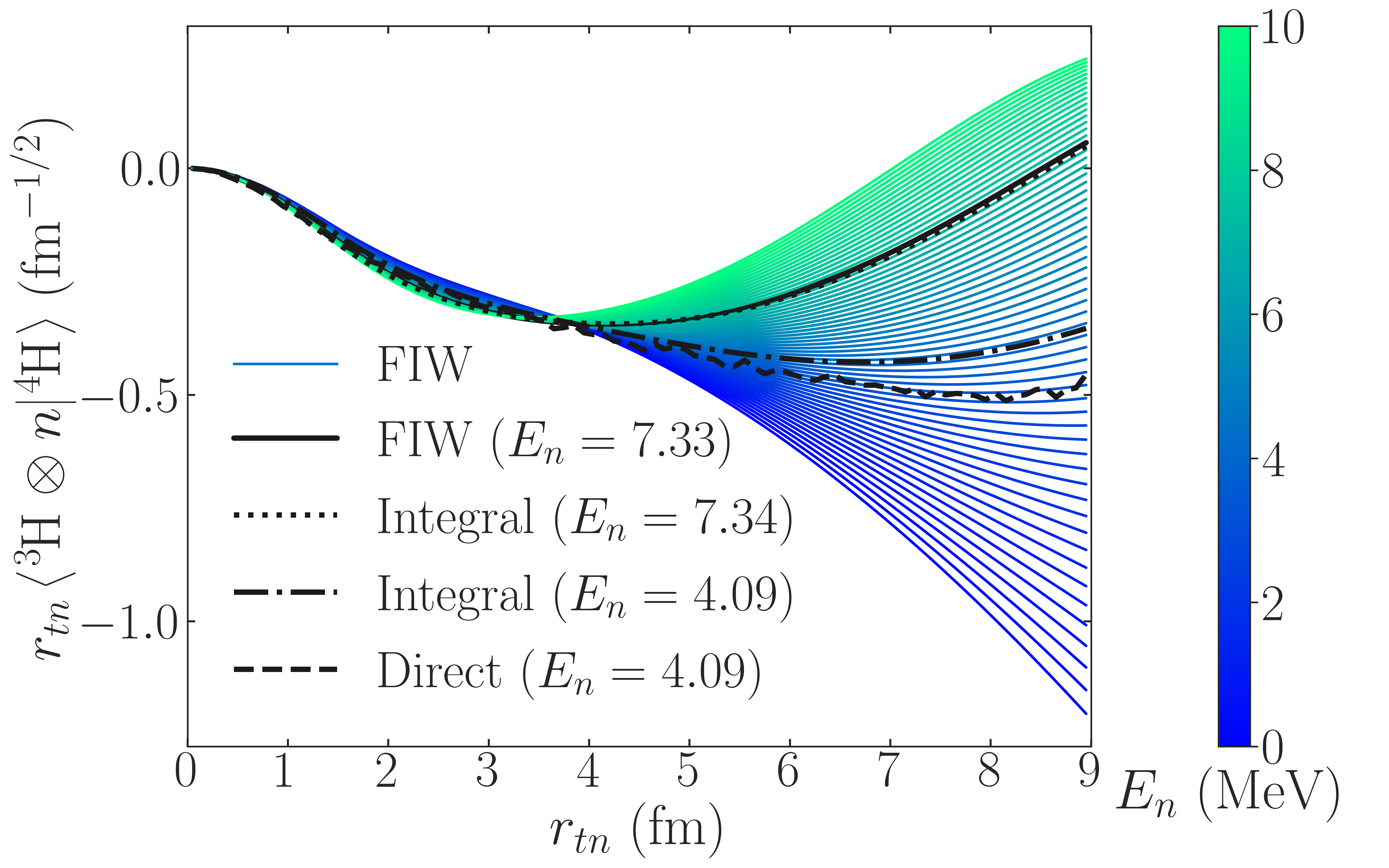}
\caption{(Color online) Overlap functions computed from a representative $\text{}^3P_{0}$ wave function with the AV18 interaction.  These have been computed using the direct method Eq. (\ref{directR}) (short-dashed, with ``shot noise''), the integral method of Eq. \ref{intR} (dash-dotted), and the fixed interior wave (FIW) approximation  (solid, energies distinguished by color or shading).  All curves except the dotted are computed from the same wave function with VMC energy corresponding to $E_n =4.09$ MeV; the dotted curve shows the integral-method overlap (analogous to the dot-dashed curve) for a separate wave function with VMC energy corresponding to $E_n=7.34$ MeV (rescaled to account for differences in normalization that arise from how the wave function fills the box). It is nearly identical to the FIW result at 7.33 MeV using the 4.09 MeV wave function, shown as a solid black line for visibility.}\label{solo_overlap}
\end{figure}

In Fig.~\ref{solo_overlap} we focus on results of these calculations for a representative state in the $^3P_0$ scattering channel.  After choosing the $\zeta$ boundary condition for this calculation, the center-of-mass energy $E_{tn}$ was computed to be 3.07 MeV ($E_n=4.09$ MeV).  The dashed curve shows the direct overlap (Eq.~(\ref{directR})), while the dash-dotted curve shows the integral-method overlap (Eq.~(\ref{intR})).  Each of these curves is shown multiplied by $r_{tn}$ to remove a trivial source of radial dependence and give functions similar to solutions of a radial Schr\"odinger equation (e.g., Eq.~(\ref{coulombscrhodinger})).  Since the direct overlap at each radius is computed only from the Monte Carlo samples that fall into a thin radial bin, the dashed curve displays  ``shot noise'' that is visible as small fluctuations with radius.  The integral-method curve, on the other hand, is smooth because at each $r_{tn}$ it contains contributions from all samples with larger $r_{tn}$; there are many more samples involved and also many shared samples contributing to any two neighboring points on the curve.

\begin{figure*}
\includegraphics[width=17.2cm]{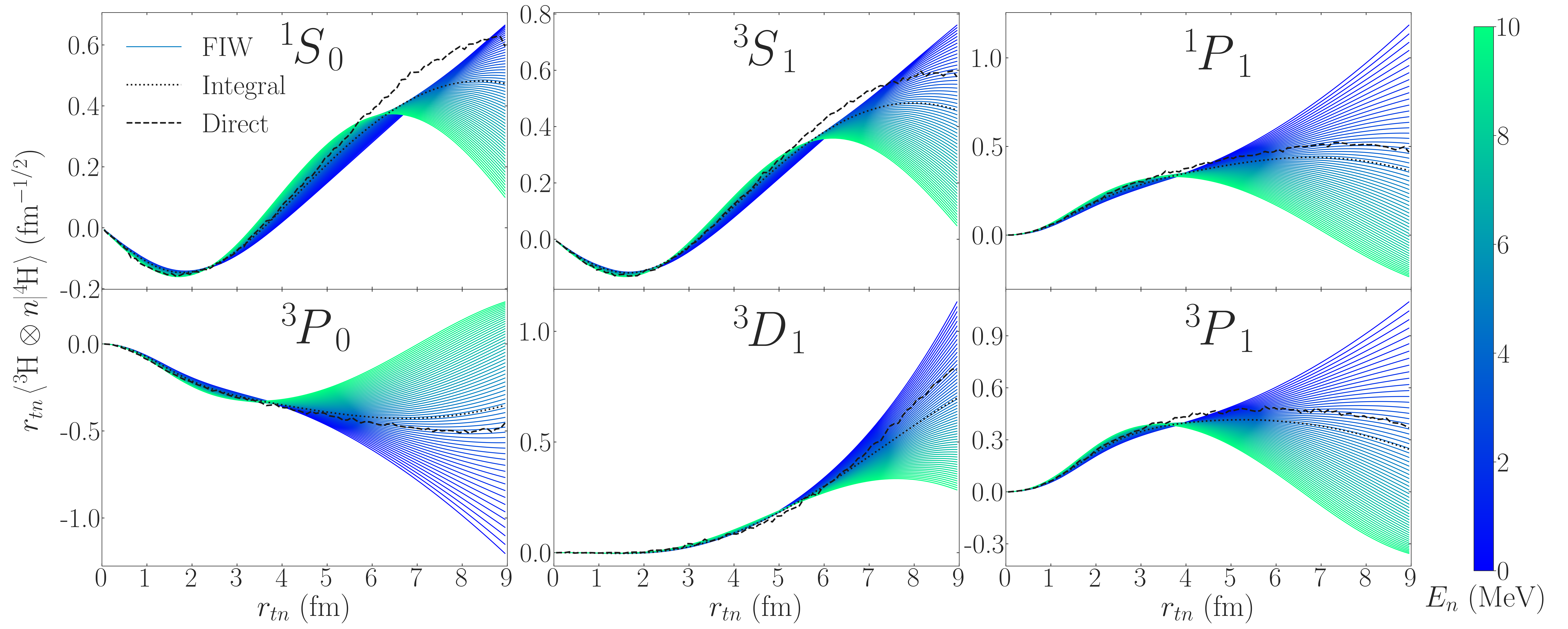}
\caption{(Color online) Overlap functions computed with AV18 for various partial waves using the same methods and labels as Fig.~\ref{solo_overlap}. The qualitative matching in the interior range is present for each channel between the direct and integral (dotted) overlaps.}\label{v18_overlaps} 
\end{figure*}

The colored (online) or shaded curves in Fig.~\ref{solo_overlap} demonstrate application of the fixed interior wave (FIW) approximation to overlap functions.  For each of those curves, we take the single optimized VMC wave function that produced the short-dashed and dot-dashed curves, but we use a different input energy for the integral relations.   From the single variational wave function, this method generates approximate overlap functions over the entire low-energy spectrum from threshold to $E_n=10$ MeV.  (No attempt has been made to rescale the wave function to unit norm inside the box for the revised probability densities implied by the new $R_c$.)  Each of these curves is consistent with the directly-computed overlap at $r < 5$ fm.  (See below for dependence of this statement on the choice of $\gamma$.)  The dotted curve shows the integral-method overlap (not FIW) for the higher-energy variational wave function that gave $E_n=7.34$ MeV; it is nearly identical to the solid black curve, which was generated from the 4.09 MeV wave function using 7.33 MeV in the FIW approximation.  These results are not unique to the $^3P_0$ channel; Fig.~\ref{v18_overlaps} shows similar results for other partial waves.

The many curves (often visible as shaded regions rather than individual curves) in Fig.~\ref{solo_overlap} illustrate a further point concerning the use of fixed interior waves: it is a very efficient use of both computer and human resources. Computed directly from the definition, each of the 60 solid (shaded) lines in Fig. \ref{solo_overlap} would require a separate wave function with a different boundary condition and separate optimization. Instead, they have been computed from the integral relations for $R_c$ from a single wave function. By skipping the separate optimization and generation of a new Monte Carlo walk (which involves recomputing wave functions) at every energy, the amount of work has been greatly reduced, especially for coupled channels.

Agreement between the methods at $r<5$ fm is important because this is the region where the VMC wave function is most reliable.  In that region, iterated pair correlations and antisymmetrization provide structure that is not dominated by one or two elements of the variational ansatz, and it is where most of the power of the VMC method arises. The $r < 5$ fm region also typically contains the largest number of Monte Carlo samples.  We therefore expect the directly-computed overlap to be rather accurate there.  In this region the integral method should be unable to improve significantly on the directly-computed overlap, and we expect the two methods to agree there if the integral method has been successfully implemented.  In Figs.~\ref{solo_overlap} and \ref{v18_overlaps} the integral (dash-dotted black line) and direct (short-dashed) overlaps are in fact nearly identical between 0 fm and 5 fm.

The choice of regularizer for $G_l$ affects both the efficiency and the accuracy of scattering calculations \cite{viviani2020n+}.   We tried several possible regularizing functions but found Eq.~\ref{reg} to be by far the most successful in reproducing overlap functions.  All of the results that we show are computed with that regularizer, and it is also the one used in recent calculations with the hyperspherical harmonics method \cite{viviani2020n+}.  The results shown are computed with a specific value of the regularization parameter $\gamma$, and we now describe how it was was chosen.

\begin{figure}
\includegraphics[width=8.6cm]{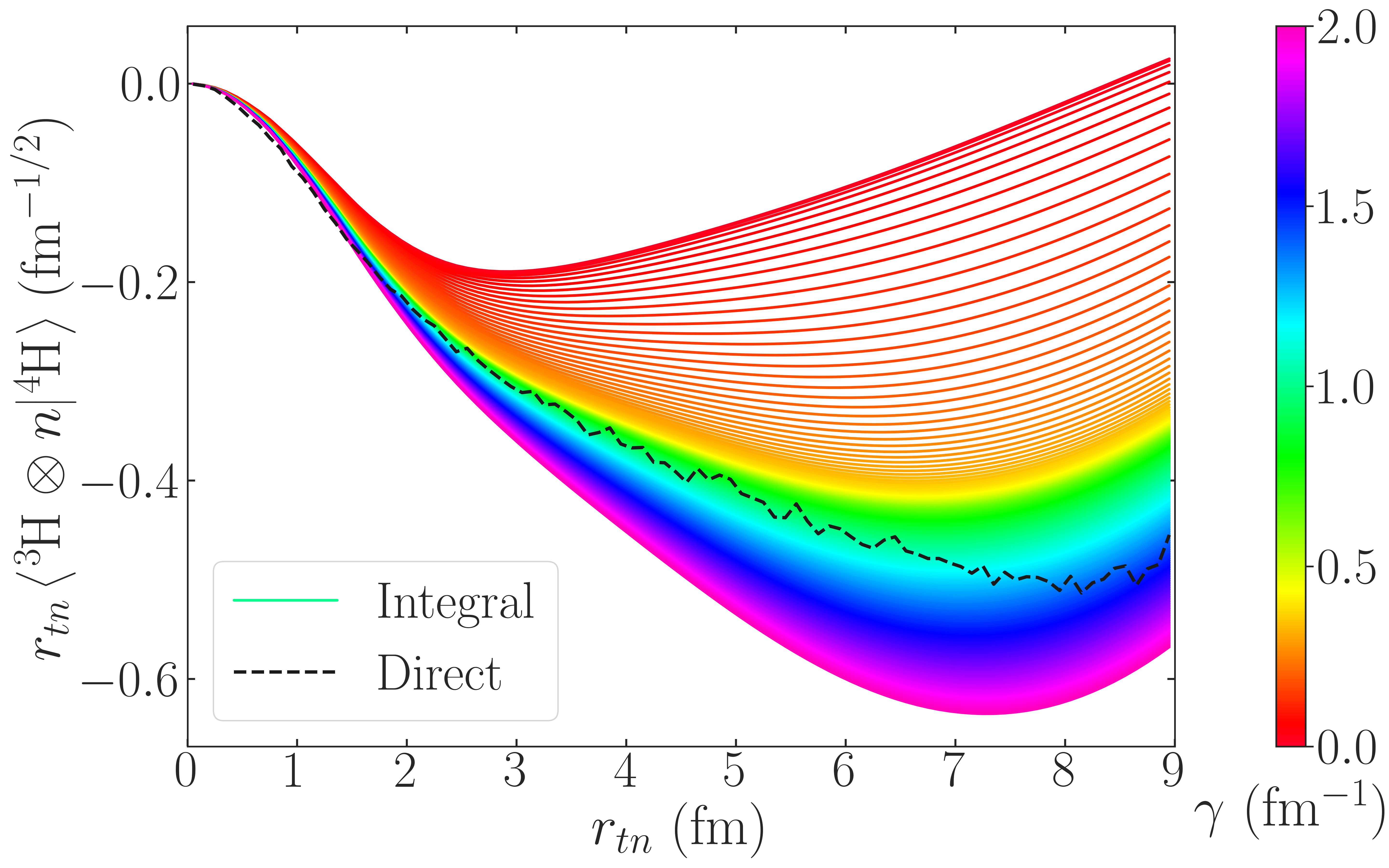}
\caption{(Color online) The effect of the regularization parameter $\gamma$ on the overlap function $r_{tn}R_c(r_{tn})$ for a $\text{}^3P_{0}$ state at $E_n=4.09$ MeV with AV18. The dashed curve shows the result of calculation from the definition, while the solid curves are the results of integral-method calculations with varying $\gamma$.  Values of $\gamma$ run from 0.01 to 2.0 fm$^{-1}$ and are indicated by color/grayscale.  Values between roughly $0.4$ and 1.4 fm$^{-1}$ produce close agreement between methods in the region where VMC is most accurate, and we choose 0.625 fm$^{-1}$ for further AV18 calculations.}\label{gamoverlaps}
\end{figure}

If $\gamma$ is chosen too small, the regularizer has effects at the box surface that violate assumptions behind Eqs.~(\ref{Aint}) and (\ref{eq:practical-a-bar}).  If it is chosen too small, the regularizer does not adequately remove effects of the singularity in Eq.~(\ref{Aexact}).
In Fig.~\ref{gamoverlaps} we vary $\gamma$  over the range 0.01 to 2.0 fm$^{-1}$ for overlap calculations of a single $\text{}^3P_{0}$ state, and we also show  the direct-method result.  Color (online) or shading indicates the value of $\gamma$, with the color/gray scale chosen to emphasize values where the methods agree.  We quantify agreement between the two methods with a sum of square errors (SSE) statistic over the $r<5$ fm region. The SSE is given by
\begin{equation}\label{SSE}
  \mathrm{SSE}_c(\gamma) \equiv \sum_{i=1}^{n_r}\left[r_i R^{D}_c(r_i)
    - r_i R^{I}_c(\gamma,r_i)\right]^2,
\end{equation}
where $R^D$ is the directly computed overlap of Eq.~\ref{directR},  $R^I$ is the integral-method overlap of Eq.~\ref{intR}, and the sum extends over the 0.1-fm-thick bins that define neutron-triton separations in the direct calculation. We continue to work with $r R_c$ rather than $R_c$, because it reduces the weight in Eq.~(\ref{SSE}) of poorly-sampled low-volume shells near $r=0$ and it reduces the severity of the divergence of $G_l$ near the origin of the Eq.~(\ref{eq:practical-a-bar}) integrand.

\begin{figure}
\includegraphics[width=8.6cm]{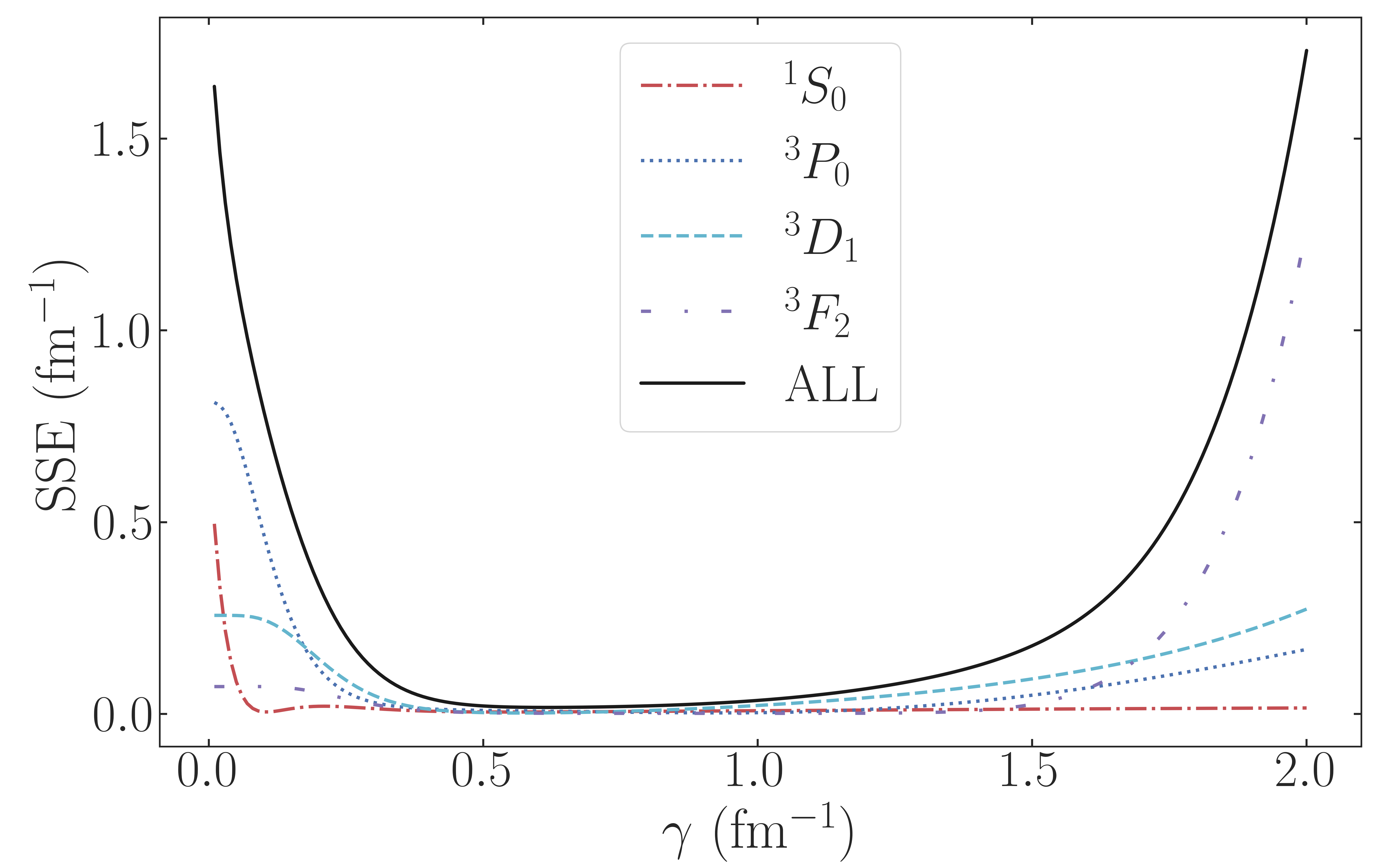}
\caption{(Color online) Comparison between direct and integral-method overlap calculations for four states in different channels near $E_n=4$ MeV at varying $\gamma$.  Agreement is quantified by the sum of square errors at $r < 5$ fm defined in Eq.~(\ref{SSE}), SSE($\gamma$), applied separately to each channel and to a sum over all four channels.  Small SSE indicates agreement between methods when $\gamma$ is small enough to regularize $G_l$ effectively at the origin but large enough not to affect the box surface.  This occurs in the flat region between 0.5 and 1.0 fm$^{-1}$.}\label{sse_fig}
\end{figure}

In principle we could choose a different gamma for each channel, but the SSE evaluation in Fig.~\ref{sse_fig} reveals a single range that works well for many channels, and we see little value in further fine-tuning for individual channels.  For AV18 we find that $\gamma = 0.625\text{ fm}^{-1}$ minimizes the SSE for all cases examined, and it lies in a ``stationary'' range of weak $\gamma$ dependence.  Repeating the analysis for the Norfolk NV2+3-Ia interaction gives a slightly smaller best $\gamma$, with the stationary range centered on $0.52\text{ fm}^{-1}$. Computing phase shifts at 4 MeV, we find that varying $\gamma$ through the range of stationary SSE only changes phase shifts by $0.5$ degree relative to the optimal $\gamma$ for both of these potentials.

\section{Scattering Results}

\label{sec:results}

Having validated our method and its implementation using overlap functions, we now turn to our main objective of computing scattering observables. We carry out these calculations using all three methods to determine asymptotic amplitudes:  direct computation of the overlap, application of the integral method at the variational energy, and use of the integral method over a range of energies with a single fixed interior wave in each channel.  Since it is customary in the recent literature on $A=4$ systems to quote laboratory energy rather than center-of-mass energy, we present most results in terms of the neutron energy when the $^3$H target is stationary,
\begin{equation}
\label{en}
    E_n = \frac{4}{3} E_\mathrm{c.m.}. 
\end{equation}
This expression neglects the small correction to the triton mass arising from its binding energy and from the difference of the proton from the neutron mass.

\subsection{Single-channel cases}
\label{sec:single-channel-cases}
\begin{figure}
\includegraphics[width=8.6cm]{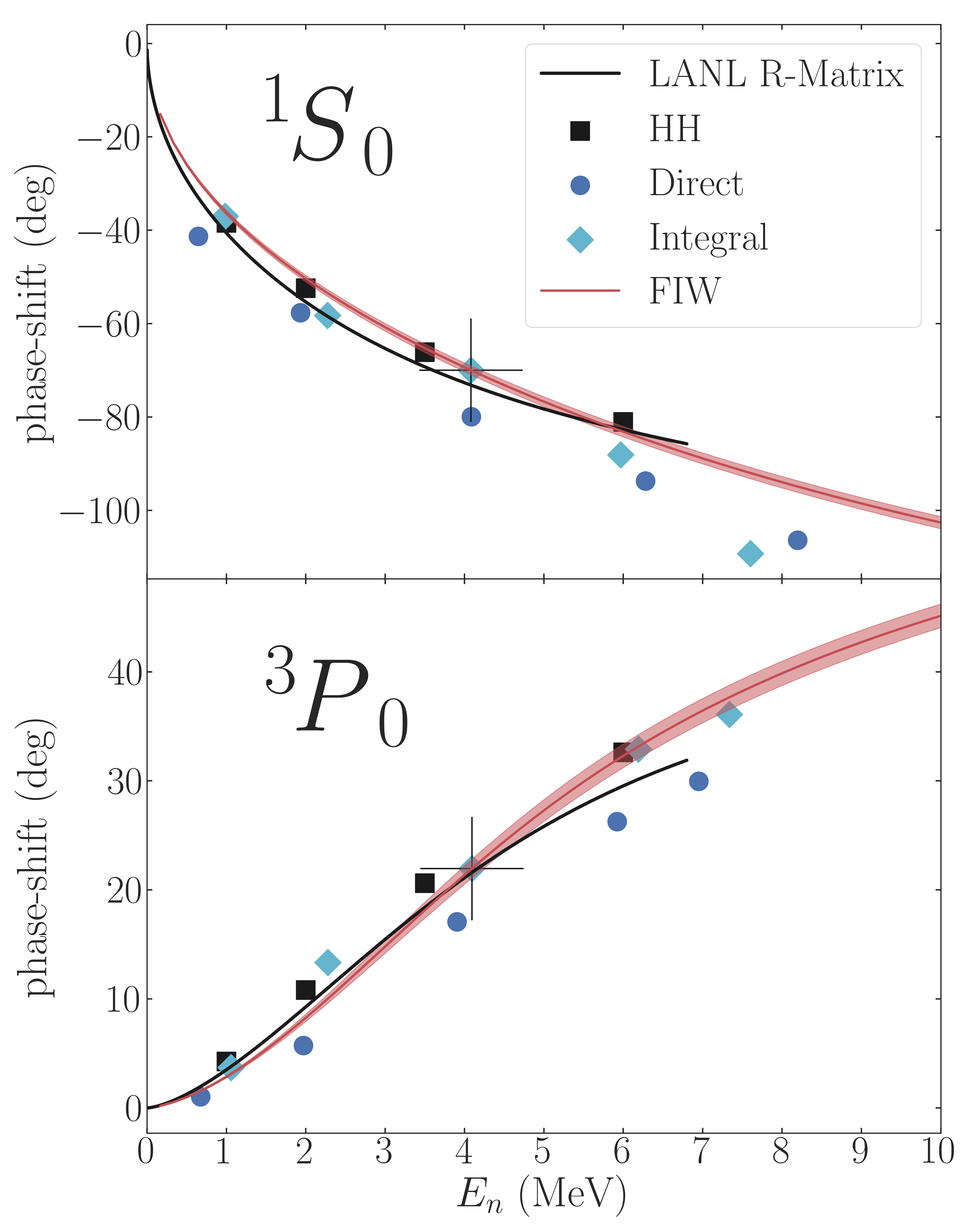}
\caption{(Color online) Single-channel phase shifts (in degrees) for $n+\text{}^3\text{H}$ with the AV18 potential, computed from VMC using the direct method (blue circles), the integral method (cyan diamonds), and fixed interior waves (red band).  The width of the band indicates the Monte Carlo sampling error.  For comparison we show an empirical  $R$-matrix model  (solid black curve labeled ``LANL R-Matrix,'' explained in the text)  \cite{hale1990neutron} and essentially exact results from the hyperspherical harmonic method \cite{viviani2011benchmark} with the same potential (black squares).  The plus symbol indicates the VMC wave function chosen to provide the fixed interior wave for computations of the red band.}\label{single_channel_phase_shifts}
\end{figure}

\begin{figure}
\includegraphics[width=8.6cm]{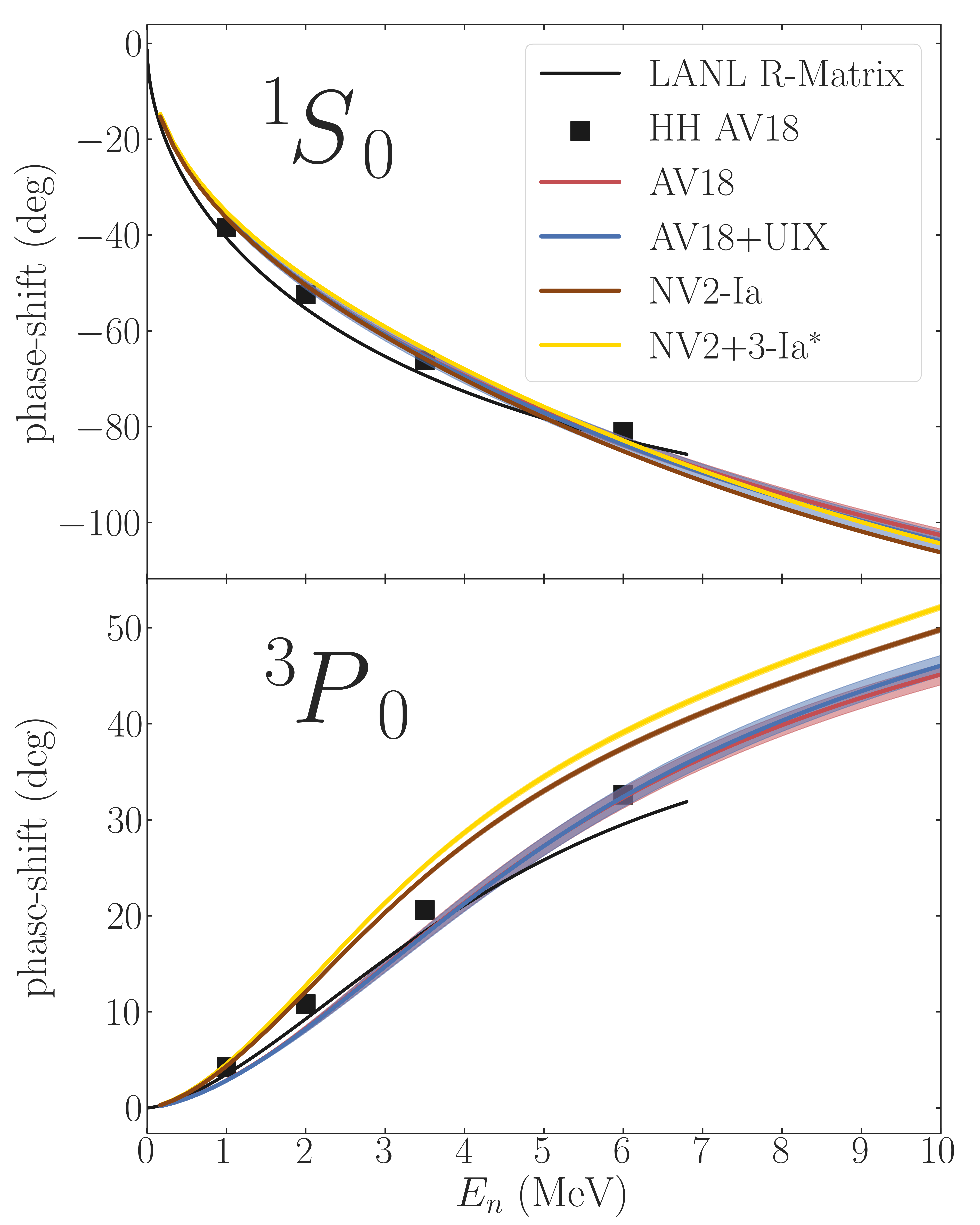}
\caption{(Color online) Single-channel phase shifts (in degrees) for $n+\text{}^3\text{H}$ in the FIW approximation for multiple interactions, with and without three-nucleon terms.  The $R$-matrix curve and AV18 benchmark calculation are as in Figs.~\ref{single_channel_phase_shifts}. The Norfolk Ia interaction both with and without the Ia$^\ast$ three-body terms shows somewhat larger attraction than the AV18-based calculations (with and without Urbana IX three-body terms) and the empirically-derived phase shifts in the $p$-waves.  Widths of the bands show Monte Carlo sampling errors.}\label{FIW_single_channel}
\end{figure}

We begin by showing the single-channel, $J^\pi = 0^+$ and $0^-$, phase shifts computed from the AV18 potential in Fig.~\ref{single_channel_phase_shifts} and from the Norfolk-Ia interactions in Fig.~\ref{FIW_single_channel}.  For comparison with our calculations, we show results for the same potential using the hyperspherical harmonics method \cite{viviani2011benchmark} as black squares.  These are well-benchmarked against other methods and can be regarded as essentially exact; any accurate calculation from AV18 should match them closely.  We also show as a solid curve empirically-derived phase shifts that were computed by fitting the much more extensive $p+\,^3\mathrm{He}$ data to a phenomenological $R$-matrix model and applying isospin symmetry \cite{hale1990neutron}.  (The isospin-rotation procedure mainly involves replacing Coulomb functions with spherical Bessel functions and shifting the level energies by a phenomenological difference of Coulomb interaction energies between the two systems.)

The direct method of computing phase shifts from VMC (circles in Fig.~\ref{single_channel_phase_shifts}) reproduces qualitative features of the phase shifts for both of these partial waves.  However, the phase shift at fixed energy comes out too low by typically $10^\circ$  for $s$-wave and $5^\circ$ for $p$-wave states.  The variational principle implies that the computed energy is higher than the true energy for any boundary condition $\zeta$, so Eq.~(\ref{tandSC}) in general gives curves that can be viewed as being either too low in phase shift or too high in energy \cite{carlson1987microscopic}.  Failure to match the exact result reflects limitations of the variational ansatz (or else failure to optimize it well).  Up to now, the main option to improve on the direct calculations from a VMC wave function has been to use the VMC wave function as the starting point for a GFMC calculation \cite{nollett2007quantum,lynn2016Chiral}; GFMC then finds the correct energy for the given boundary condition more accurately before computation of the phase shift.

The diamond shapes in Fig.~\ref{single_channel_phase_shifts} show results of applying the integral relations in Eqs.~(\ref{Bint}) and (\ref{Aint}) to VMC wave functions and using the single-channel relation $\tan\delta_c = B_c/A_c$.  The energy assumed in the integral relations is equal to the variational energy of each state.  These results are in much closer alignment with the exact results, but with some scatter away from their trend for individual VMC results.   The variation presumably reflects the quality of each wave function optimization.  The energy used in computing both the  circles and the diamonds is measured relative to a neutron and triton at rest infinitely far apart.  For this we take the difference of computed VMC energies between the scattering state and our best VMC triton wave function (rather than, e.g., the exact triton energy for AV18).

Finally, in each partial wave we apply the FIW approximation by choosing one VMC solution (marked in the graphs with a cross) and computing integral-relation phase shifts at all energies from that single wave function.  The result is shown as the red curve with shaded band corresponding to Monte Carlo statistical errors on the integrals.  By its construction the red curve passes through the point with the cross on it, which was chosen for its location in the middle of the energy range of interest.  Table \ref{table:nt-shifts} shows that the results are within $2.5^\circ$ of exact phase shifts for AV18.   Inspection of $^3P_0$ phase shifts in Fig.~\ref{single_channel_phase_shifts} suggests that better results might have been obtained by choosing the wave function for FIW treatment based on how well it matches exact results; even without exact results for comparison,  a wave function with outlying low phase shifts relative to other VMC points could be avoided as possibly poorly-optimized.   Nonetheless, deviation of the $p$-wave phase shifts from the trend of both the $R$-matrix model and the benchmark indicates that the FIW loses accuracy when it proceeds too far above the energy of the wave function used; this appears to happen around 6 MeV in the present case.

These results establish the integral method and its variant with fixed interior waves as useful ways to obtain approximate phase shifts from VMC wave functions.  The combination of integral relations with VMC wave functions evidently does not achieve the precision available with other computational frameworks in the $n+\,^3\mathrm{H}$ system.  However, it greatly improves both the quality and the efficiency of scattering calculations possible using VMC, which might be more readily applied to larger systems than the other solution methods.  More importantly, the close relationship between VMC and GFMC should allow what we have developed here to be taken over to GFMC. GFMC produces much more precise wave functions than VMC but still suffers from subtle difficulties in computing the outer parts of particle-in-a-box solutions at high precision and in generating sufficiently exact energies to compute observables near threshold with the scattering formalism currently in use \cite{nollett2007quantum}.

\subsection{Coupled channels}
\label{sec:coupled-channels}
\begin{figure*}
\includegraphics[width=17.2cm]{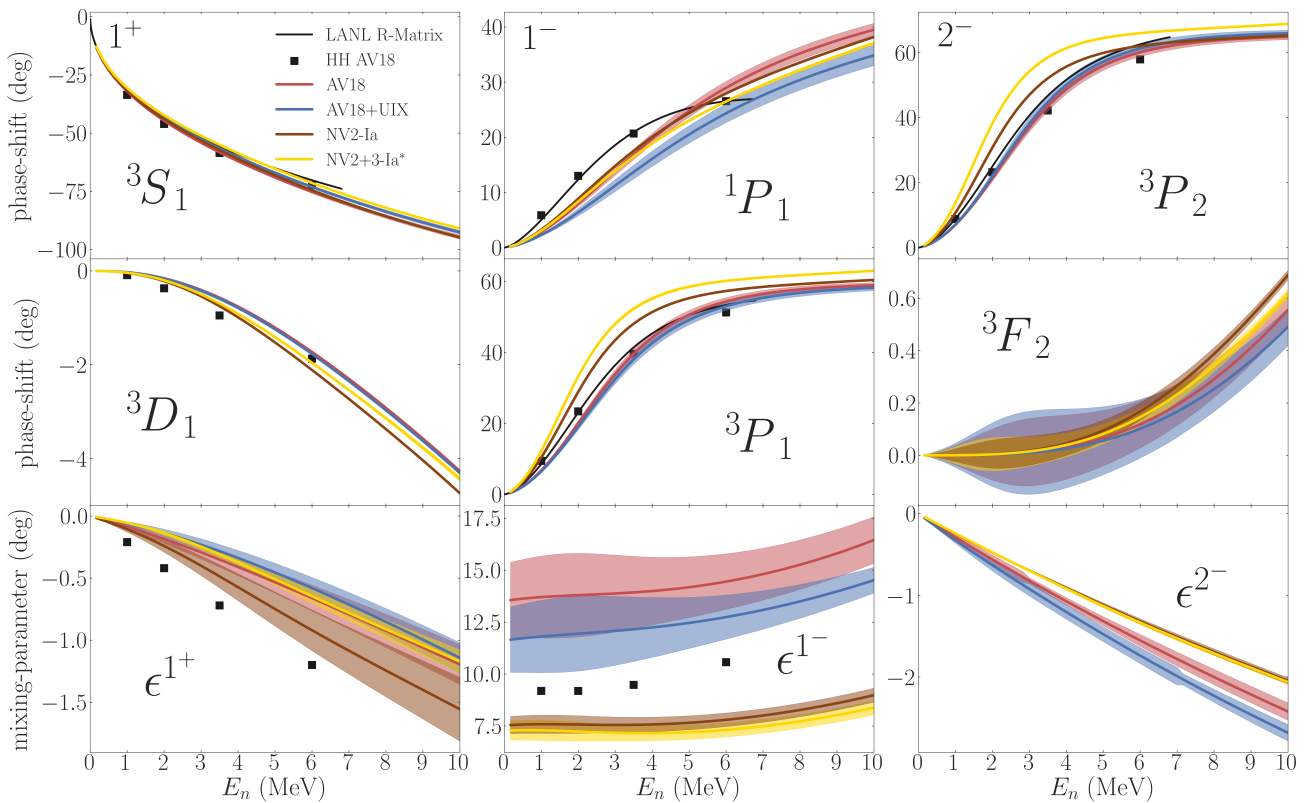}
\caption{(Color online)  Phase shifts and mixing parameters (in degrees) for $n+\text{}^3\text{H}$ in the FIW approximation with the same interactions as Fig.~\ref{FIW_single_channel}. The $R$-matrix curve and AV18 benchmark calculation are as in Figs.~\ref{single_channel_phase_shifts} and \ref{FIW_single_channel}.  The AV18 phase shifts are within $2.5^\circ$ of the benchmark, except for an apparent difficulty shared by calculations with all potentials in the $^1P_1$ channel. Widths of the bands show Monte Carlo sampling errors.}\label{FIW_phase_shifts}
\end{figure*}

Proceeding to other partial waves in Fig.~\ref{FIW_phase_shifts}, we show only results with fixed interior waves, always generated from one VMC wave function per channel near 4 MeV.  The remaining partial waves relevant at low energy come in three sets of coupled channels, though the channel coupling is extremely weak in the $1^+$ and $2^-$ states (where it arises from the tensor force); the coupling is somewhat stronger between the two spin combinations in the $p$-wave states with $J^\pi=1^-$.  Phase shifts and the largest mixing parameter from Fig.~\ref{FIW_phase_shifts} are given in Table \ref{table:nt-shifts} at the energies of published benchmark calculations \cite{viviani2011benchmark}.  

\begin{table*}[hp]
  \caption{$n+\text{}^3\text{H}$ phase shifts and mixing parameters computed using the FIW approximation for various interaction models.  The hyperspherical harmonics (HH) results are from Ref.~\cite{viviani2011benchmark}.}
  \label{table:nt-shifts}
      \begin{tabular}{l c c c c c c c c } 
      \hline\hline
        & $^1S_0$ & $^3P_0$ & $^3S_1$ & $^3D_1$ & $^3P_2$ & $^1P_1$ & $^3P_1$ & $\epsilon^{1-}$\Tstrut\Bstrut\\
        \hline
        $1.0$ MeV\\
        \hline
        HH AV18 & $-$38.44 & 4.26 & $-$33.57 & $-$0.09 & 8.82 &  5.87 & 9.44 & 9.19\\
        \hline
        AV18 & $-$36.3(5) & 2.9(1) & $-$32.1(3) & $-$0.03(2) & 7.0(3) & 2.8(2) & 6.8(3) & 13.7(19)\\
        AV18+UIX & $-$36.2(8) & 2.8(1) & $-$31.1(2) & $-$0.03(3) & 7.1(4) & 2.3(2) & 6.5(3) & 11.8(17)\\
        NV2-Ia & $-$36.40(9) & 4.34(5) & $-$31.71(6) & $-$0.04(2) & 10.6(1) & 3.24(4) & 10.4(1) & 7.6(4)\\
        NV2+3-Ia$^{\ast}$ & $-$35.12(9) & 4.57(6) & $-$30.58(7) & $-$0.04(1)& 12.2(2) & 3.08(5) & 13.7(2) & 7.3(5)\\
        \hline
        $2.0$ MeV\\
        \hline
        HH AV18 & $-$52.41 & 10.82 & $-$46.04 & $-$0.37 & 23.21 & 13.00 & 23.39 & 9.19\\
        \hline
        AV18 & $-$50.3(7) & 8.3(3) & $-$44.7(4) & $-$0.16(2) & 21.3(10) & 7.95(51) & 20.0(8)  & 13.8(20)\\
        AV18+UIX & $-$50.3(11) & 8.2(3) & $-$43.3(3) & $-$0.17(4) & 21.9(12)  & 6.4(7) & 19.0(8) & 11.9(18)\\
        NV2-Ia & $-$50.5(1) & 12.1(1) & $-$43.92(8) & $-$0.22(2) & 30.8(3) & 8.86(11) & 28.9(3)  & 7.6(4)\\
        NV2+3-Ia$^{\ast}$ & $-$48.7(1) & 12.8(2) & $-$42.3(1) & $-$0.20(1) & 38.1(5) & 8.35(12) & 33.3(5)  & 7.2(5)\\
        \hline
        $3.5$ MeV\\
        \hline
        HH AV18 & $-$66.14 & 20.61 & $-$58.53 & $-$0.95 & 42.22 & 20.68 & 39.63 & 9.48\\
        \hline
        AV18 & $-$65.3(9) & 18.1(7) & $-$58.3(5) & $-$0.57(2)& 43.6(17) & 16.9(9) & 39.6(13)  & 13.9(17)\\
        AV18+UIX & $-$65.4(14) & 18.1(7) & $-$56.5(5) & $-$0.59(4) & 44.8(19) & 13.7(12) & 38.0(14)  & 12.2(16)\\
        NV2-Ia & $-$65.8(2) & 24.1(3) & $-$57.3(1) & $-$0.75(2) & 52.1(3) & 17.5(2) & 48.1(4)  & 7.6(4)\\
        NV2+3-Ia$^{\ast}$ & $-$63.7(2) & 25.2(3) & $-$55.0(1) & $-$0.69(1) & 58.6(4) & 16.5(2) & 52.3(5) & 7.2(4)\\
        \hline
        $6.0$ MeV\\
        \hline
        HH AV18 & $-$81.05 & 32.61 & $-$72.40 & $-$1.87 & 57.94 & 26.55 & 51.27 & 10.57\\
        \hline
        AV18 & $-$83.2(11) & 32.3(11) & $-$75.2(7) & $-$1.73(1) & 60.1(13) & 29.0(12) & 54.4(11)  & 14.4(13)\\
        AV18+UIX & $-$83.7(17) & 32.4(11) & $-$73.1(6) & $-$1.77(2) & 61.3(15) & 24.3(17) & 53.0(12) & 12.7(11)\\
        NV2-Ia & $-$85.1(2) & 37.5(3) & $-$74.5(2) & $-$2.11(2) & 62.1(2) & 27.9(2) & 57.3(3) & 7.8(3)\\
        NV2+3-Ia$^{\ast}$ & $-$82.9(3) & 39.1(3) & $-$71.4(2) & $-$1.96(2) & 66.0(3) & 26.5(4) & 60.2(3) & 7.3(3)\Bstrut\\
        \hline\hline
      \end{tabular}
\end{table*}

Considering only AV18 for the moment, we find good agreement with the benchmark for all of the phase shifts except in the $^1P_1$ channel, where the curve is qualitatively different from both the benchmark and the empirically-derived curve.  We obtain very similar results when we carry out the calculation with other potentials, suggesting some systematic problem with  our calculations in this specific channel.  After extensive checking, we have been unable to find a coding error, and we suggest tentatively that it may be a shortcoming of the variational ansatz specific to this configuration.  It will be informative in the near future to see if the problem persists in GFMC calculations that use the VMC solutions here as starting points.

The $1^+$ mixing parameter is very small in both our AV18 calculation and the hyperspherical harmonics calculation, smaller than $2.5^\circ$.  In that sense there is good agreement, though the VMC results are about half the size of the hyperspherical result.  For $1^-$ states the VMC mixing parameter comes out somewhat larger (roughly $14^\circ$ instead of $10^\circ$ over the whole energy range) but they are of similar magnitude, and one of channels here is the $^1P_1$ channel where something has apparently gone wrong in our VMC calculations as discussed above.

\subsection{Three-nucleon and chiral potentials}

We also carried out calculations that combine the AV18 two-body potential with the Urbana IX (UIX) three-nucleon interaction \cite{Pudliner1995Quantum,Carlson2015QMC}, and that use the Norfolk family of local chiral potentials \cite{piarulli2018light,baroni2018local}.  Results for AV18+UIX and for the NV1a and NV2+3-Ia$^\ast$ potentials are shown alongside the AV18-only results in Figs.~\ref{FIW_single_channel}  and \ref{FIW_phase_shifts} and in Table \ref{table:nt-shifts}.  These two Norfolk potentials differ in their inclusion or not of three-nucleon terms and in what data were used in fitting them; NV2-Ia consists only of two-body terms, while NV2+3-Ia$^\ast$ includes a three-body interaction.  It is evident from the phase shift graphs that while results in other channels are very similar between potentials, the Norfolk interactions provide somewhat larger attraction in the $p$-wave channels than appears to be supported by the isospin-symmetry-based $R$-matrix curve.  No strong qualitative dependence on three-body terms is evident in the cases shown, though there is some difference.   Qualitative agreement of all calculations in the $s$-wave channels is not surprising, since $s$-wave scattering in this and many light systems has the character of scattering from a hard sphere, due to antisymmetry constraints on the wave function \cite{nollett2007quantum,aurdal70}.

\begin{figure*}
\includegraphics[width=17.2cm]{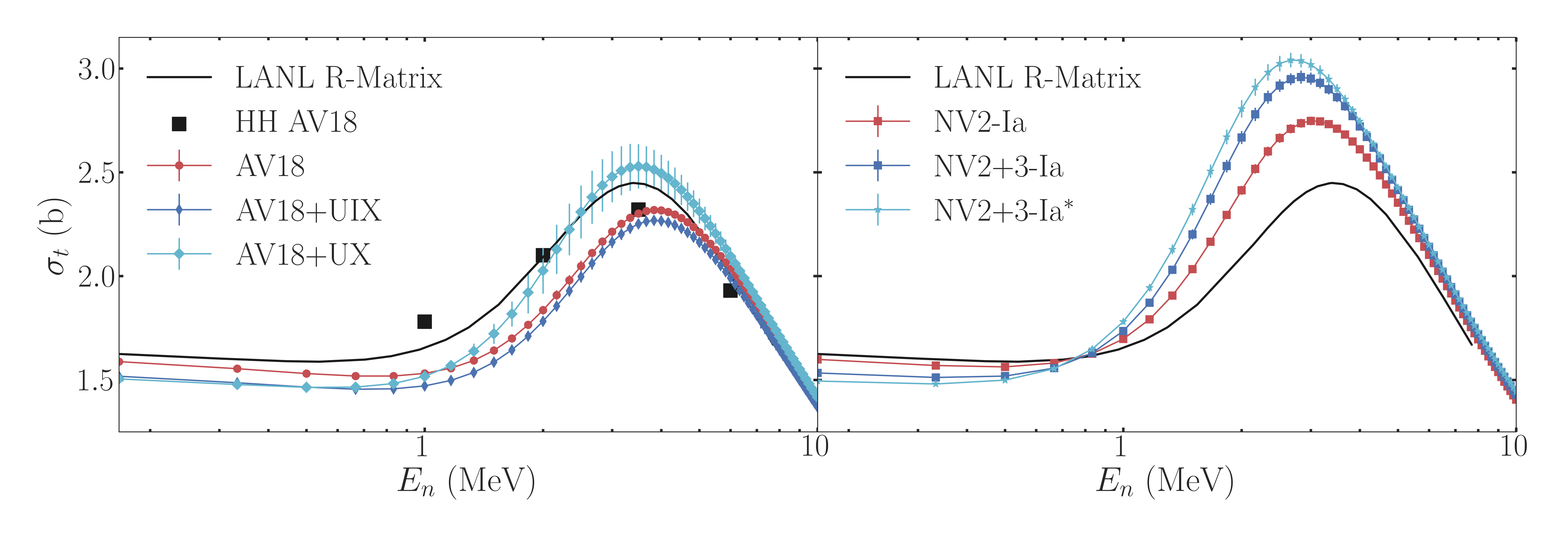}
\caption{(Color online) $n+\text{}^3\text{H}$ total cross section (in barns) as a function of neutron energy for various interactions.  The left panel shows results for AV18 with and without the Urbana IX or Urbana X three-body terms. The right panel shows the Norfolk Ia potential with and without the Ia or Ia$^\ast$ three-body terms; error bars show Monte Carlo statistical uncertainties. Other symbols are as in Fig.~\ref{single_channel_phase_shifts} except that the LANL total cross section is from Ref.~\cite{BrownENDF2018}. In general, the Norfolk Ia interactions give somewhat too-pronounced resonance structures in $p$-waves, while $s$-wave scattering correlates mainly with triton binding energy regardless of potential.}\label{tcs}
\end{figure*}

We show total cross sections for these interactions and for two others in Fig.~\ref{tcs}.  In the left panel, we show results for AV18 alone and for AV18+UIX; we also show results for AV18 with the Urbana X three-nucleon interaction, which has a very similar structure to Urbana IX but has been tuned to produce binding energies closer to those of the more computationally expensive Illinois-7 interaction \cite{crespo2020}.  Although AV18+UIX produces very similar results to AV18 alone here, it is evident that a different choice of three-nucleon interaction can have a noticeable effect on the $p$-wave peak around 3 MeV.  The right panel shows total cross sections for NV2+3-Ia$^\ast$ and for NV2+3-Ia, which have three-body terms tuned to match differing input data.  Here the choice of three-body interaction also has a significant effect on the resonance structure in the $p$-waves.  In general, AV18 and AV18+UIX underpredict the strength and width of the resonance feature while all of the Norfolk interactions (not just those shown here) overpredict them.

We summarize our $s$-wave calculations for a large collection of potentials by presenting total cross sections $\sigma_t$ for thermal neutrons and coherent scattering lengths $a_c$ in Table \ref{table1}.  These are computed from the singlet and triplet $s$-wave phase shifts via
\begin{equation}
\label{thermalcs}
 \sigma_t =  \frac{\pi}{k^2}(\sin^2\delta_0 + 3\sin^2\delta_1)  
\end{equation}
and
\begin{equation}
\label{ac}
 a_c =  \frac{1}{4k}(\sin\delta_0 + 3\sin\delta_1),
\end{equation}
applying integral relations with $E_{c.m.} = 0.025$ eV to the usual fixed interior wave in each channel with variational energy $E_n\simeq 4$ MeV.  We found no significant evolution of $\sigma_t(E)$ below 10 eV.  As indicated in Table \ref{table1}, the total cross section has been measured to below 50 keV and extrapolated to thermal energies yielding a result of \mbox{$\sigma_t=1.70\pm0.03$ b} \cite{phillips1980neutron}. The two most recent measurements of the experimental coherent scattering length  \cite{hammerschmied1981measurements,rauch1985re} were carried out by the same group, with a more advanced setup for the second measurement.  A third empirically-derived coherent scattering length comes from essentially the same  Coulomb-corrected $R$-matrix calculation shown in our graphs \cite{hale1990neutron}. 

\begin{table}[h]
  \caption{$n+\text{}^3\text{H}$ Thermal-neutron cross section $\sigma_t$ (in barns), coherent scattering length $a_c$ (in fm), and $\text{}^3$H binding energy (in MeV) computed in VMC with various interactions. The VMC results are computed in FIW approximation at $E_{c.m.} = 0.025$ eV using wave functions of variational energy $E_n\approx 4$ MeV. For comparison we show values from the  essentially exact hyperspherical harmonic (HH) method for AV18 and AV18+UIX, for the Coulomb-corrected $R$-matrix, and from experiment. The VMC binding energies here are not the true binding energies for each given interaction, but instead are our best-optimized VMC wave binding energies, which establish the threshold energy for present purposes. }
  \begin{tabular}{l c c c}
    \hline\hline
    Interaction & $\sigma_t$& $a_c$& $B_3$\Tstrut\\
    \hline
    AV18 &1.632(12)&3.598(27)&7.484(2)\\
    AV18+UIX &1.558(13)&3.513(29) &8.277(2)\\
    AV18+UX &1.543(15)&3.496(32)&8.254(6)\\
    NV2-Ia &1.648(3)&3.615(6)& 7.602(9)\\
    NV2-Ib &1.656(3)&3.622(7)&7.339(9)\\
    NV2-IIa &1.614(4)&3.579(9)&7.715(5)\\
    NV2-IIb &1.734(68)&3.71(15)&7.646(14)\\
    NV2+3-Ia &1.579(5)&3.535(10)&8.179(9)\\
    NV2+3-Ib &1.558(4)&3.515(8)&8.170(15)\\
    NV2+3-IIa &1.539(3)&3.494(6)&8.193(10)\\
    NV2+3-IIb &1.566(8)&3.522(18)&8.236(14)\\
    NV2+3-Ia$^\ast$ & 1.536(3)&3.490(7)&8.205(8)\\
    NV2+3-Ib$^\ast$ &1.580(5)&3.538(11)&8.161(14)\\
    NV2+3-IIa$^\ast$ & 1.544(5)&3.498(10)&8.218(17)\\
    NV2+3-IIb$^\ast$ &1.557(5)&3.513(10)&8.212(22)\\
    \hline
    HH \cite{viviani2020n+,marcucci2020hyperspherical}\\
    AV18 &1.85&3.83&7.624\\
    AV18+UIX &1.73&3.71&8.479\\
    \hline
    $R$-matrix \cite{hale1990neutron}& - & 3.607(17)&-\\ 
    \hline
    EXPT. & 1.70(3) \cite{phillips1980neutron} & 3.82(7) \cite{hammerschmied1981measurements}& 8.475 \cite{wang2012ame} \\
    &  & 3.59(2) \cite{rauch1985re}\Bstrut\\
    \hline\hline
  \end{tabular} \label{table1}
\end{table}

Overall, the results in Table \ref{table1} depend only weakly on potential.   It is known that the triton binding energy $B_3$ in a given model correlates with the $s$-wave phase shifts, analogously to the ``Phillips line'' that correlates $B_3$ with the neutron-deuteron scattering length when potentials are varied \cite{kirscher2013}.  The VMC results in Table \ref{table1}  support a reliable negative-slope correlation between $B_3$ and $a_c$, with $B_3$ taken not as the exact value for the potential but as our best-optimized variational result.  The hyperspherical harmonics results for AV18 with and without Urbana IX suggest a correlation that has the same slope but is offset by about 0.2 fm in scattering length, presumably reflecting the same difference in precision between VMC/FIW and exact results seen above.  Considering the offset from the hyperspherical harmonics calculations, it would appear that in general values for a given potential can be estimated by adding about 0.2 fm to $a_c$ or \mbox{0.2 b} to $\sigma_t$ computed from VMC/FIW.  Essentially all of the differences among the Norfolk potentials in Table \ref{table1} are attributable to the correlation between the VMC-optimized (not exact) $B_3$ and $a_c$.

\section{Conclusion}
\label{sec:conclusion}

We have explored the application of integral relations to calculations of phase shifts and scattering observables in a quantum Monte Carlo context.  This approach replaces operations on the most poorly-computed parts of a quantum Monte Carlo wave function with integrals over the well-computed interaction region of the wave function.  Although we expect the utility of the integral relations to be mainly in heavier systems and in Green's function Monte Carlo calculations, we have developed the method by applying it to neutron-triton scattering using variational Monte Carlo wave functions.  This allows us to compare our results against well-benchmarked and essentially exact results from the recent literature for the same nucleon-nucleon interaction.  It also minimizes turnaround time for computation in this exploratory work and allows better opportunity to distinguish wave function problems from integral-relation problems in code development than if we had jumped directly to GFMC.

As an important test of the method, we began with an examination of the spectroscopic overlaps of full four-body wave functions onto $n+\,^3\mathrm{H}$ configurations.  This examination revealed that the effect of the integral relations can be viewed largely as identifying the correct outer parts of the wave function for consistency with the more accurately-computed interior parts of the wave function; accurate calculation of the outer regions is known to be difficult for both VMC and GFMC (for different reasons in each case and not as badly for GFMC).  The results of  overlap calculations also indicated the feasibility of using a single variational wave function for calculation of overlaps and scattering observables over a whole range of energies.  This works because the short-range part of the wave function is insensitive to changes in the total energy that are small relative to the potential energy.  We call this approach the fixed interior wave or FIW approximation.

Having tested the integral relations through their application to overlaps, we then applied them to the computation of phase shifts for the AV18 nucleon-nucleon interaction.  Here they give corrected surface amplitudes of particle-in-a-box wave functions, and those amplitudes are used to infer phase shifts and channel-mixing parameters.  The results represent a considerable improvement over na\"ive examination of the box surface, especially for scattering states with coupled channels, where a matrix inversion is needed to compute observables.  Moreover, we found that the fixed interior wave approximation allows calculation of phase shifts within a couple of degrees of exact values for the AV18 potential, except in the case of singlet $p$-waves.  For that case, we see what appears to be a similar shortcoming in all of the potentials that we used, so we conjecture either a fundamental shortcoming of our variational ansatz or possibly a coding error that escaped extensive testing.

We then presented calculations that combine the Urbana IX or Urbana X three-body interaction with AV18, as well as calculations with several variants of the Norfolk local chiral interactions.   These represent some of the first calculations of nucleon-nucleus scattering with the Norfolk interactions.  We found them to have generally stronger resonance features in $p$-wave scattering than AV18; their neutron scattering lengths correlate strongly with the (VMC-computed) triton binding energy and otherwise seem consistent with the AV18 and AV18+UIX/X results.

As mentioned above, the present work is intended as a step in technique development, not the final destination.  In combination with VMC, the integral method produces considerably more precise phase shifts than previously seemed possible with the standard nuclear VMC ansatz, and for substantially less investment of resources into optimizing multiple wave functions.  This makes VMC immediately more useful for approximate calculations of nucleon-nucleus scattering in the $A\leq 10$ range, and there is no reason why it could not also be applied to alpha-nucleus scattering in at least cases (like the $^7$Be and $^7$Li systems) where alpha clustering is already a prominent part of the variational ansatz.  The integral method also removes what appeared to be a serious obstacle in determining surface amplitudes well enough for coupled-channel calculations to be at all worthwhile using VMC.  None of this is surprising, since the integral relations have provided a crucial tool in hyperspherical harmonics calculations for about a decade \cite{kievsky2010variational,Romero2011General,viviani2020n+,marcucci2020hyperspherical}, but development specific to the quantum Monte Carlo context was needed, and the amount of benefit to be gained from application to VMC wave functions in particular was unknown.

We expect the main payoff of this work to lie in its future application to GFMC wave functions.  The outer parts of diffuse wave functions, including the particle-in-a-box wave functions used for scattering, converge slowly in GFMC.  Even though the results are considerably more accurate than VMC, substantial attention to these issues has been needed to obtain results of acceptable precision in past scattering calculations \cite{nollett2007quantum} as well as some bound-state calculations \cite{lu2013}.  Since the GFMC wave functions are more accurate than VMC wave functions (even in their outer regions), the amount by which the surface amplitudes have to be ``corrected'' by the integral relations will be considerably smaller.  It is our hope that the integral relations will enable high-quality GFMC scattering results with considerably reduced human effort, and make treatment of coupled-channels problems feasible.  At least for nucleon-nucleus scattering, implementation of integral relations in GFMC will be closely related to work already done on spectroscopic overlaps \cite{brida2011quantum}, and it will use the same computer routines developed here.  

\appendix
\section{Scattering matrices and parameters}
\label{appendix:scattering}

In two-cluster scattering there are three common representations of the asymptotic wave function.  In these,  wave function amplitudes at infinity are related to each other by the $T-$, $S-$, or $K-$matrix, depending on whether the wave function is represented by plane, spherical, and/or standing waves.  Each formalism provides a natural way to view some part of our calculations.

In the $K$-matrix formalism that we use for many-body calculations the wave function at large distance is written as in Eq.~(\ref{Kaysmptote}).  Alternatively, the same function can be written in terms of incoming plane-wave components and outgoing spherical waves $\mathcal{H}_c^{+} = \mathcal{G}_c + i\mathcal{F}_c$ to obtain the $T$-matrix formalism, where schematically
\begin{equation}
  \label{Taysmptote}
  \Psi(\text{all } r_c \to \infty)
  = \sum_c \left(\mathscr{A}_c\mathcal{F}_c + \mathscr{B}_c\mathcal{H}^+_c\right)
\end{equation}
and the surface amplitudes are $\mathscr{A}_c$, $\mathscr{B}_c$.  Finally, the $S$-matrix formalism is written in terms of incoming and outgoing spherical waves $\mathcal{H}_c^{\pm}$ (with $\mathcal{H}_c^{\pm}=\mathcal{G}_c \pm i\mathcal{F}_c$) so that 
\begin{equation}
  \label{Saysmptote}
  \Psi(\text{all } r_c \to \infty)
  = \sum_c \left(\alpha_c\mathcal{H}^-_c + \beta_c\mathcal{H}^+_c\right);
\end{equation}
in this case the amplitudes are $\alpha_c$ and $\beta_c$.

All of these formulations are equivalent and can be interconverted.  When each set of amplitudes is written as a vector, the relation between them is a  matrix that predicts scattering outcomes by relating incoming to outgoing amplitudes:
\begin{eqnarray}
&&\mathbf{B} = \hat{K}\mathbf{A}\label{kmat}\\&&
\mathbf{\mathscr{B}} = \hat{T}\mathbf{\mathscr{A}}\label{tmat}\\&&
\mathbf{\beta} = \hat{S}\mathbf{\alpha} \label{smat}.
\end{eqnarray}
Regardless of the form chosen, the amplitude vectors in Eqs.~(\ref{kmat})-(\ref{smat}) can in principle be read out of any wave function solution and the scattering matrices found by inverting these equations; for $N_c$ coupled channels, the inversion requires $N_c$ linearly independent solutions.  

Because particle number is conserved, single-channel scattering satisfies the constraint $|\alpha|=|\beta|$ and allows scattering matrices (actually scalars in this case) to be written in terms of a phase shift $\delta$.  Then $K = \tan\delta$, $T = e^{i\delta}\sin\delta$, and $S = e^{2i\delta}$.  In coupled-channel scattering each matrix has dimension $N_c \times N_c$.

In the $n+\,^3\mathrm{H}$ system each scattering matrix is block-diagonal and splits into $1\times 1$ and $2\times 2$ blocks with definite parity $\pi$ and total angular momentum $J$.  For comparability with the literature, we report results in terms of channels defined by orbital quantum number $L$ and by coupling neutron and triton spins to total spin quantum number $S$ (instead of the $l_c$ and $j_c$ of Eq.~\ref{12wf}). In this basis, the $2\times 2$ coupled pairs of channels either have $L = J\pm 1$ or else have $S=0,1$ together with $L=J$.   For these cases,  we write the $2\times 2$ $S$-matrix block in the Blatt-Biedenharn or eigenphase representation \cite{blatt1952} as
\begin{equation}
\label{blatt}
    \hat{S} = 
\hat{O}^T
\begin{pmatrix}
e^{2i\delta_{-}}
&0\\
0
&e^{2i\delta_{+}}
\end{pmatrix}
\hat{O},
\end{equation}
where 
\begin{equation}
\label{blattO}
    \hat{O} = 
\begin{pmatrix}
\cos\epsilon^{J\pi}
&\sin\epsilon^{J\pi}\\
-\sin\epsilon^{J\pi}
&\cos\epsilon^{J\pi}
\end{pmatrix}.
\end{equation}
In the absence of channel mixing, $\hat{O}$ is the identity matrix and the $\delta_{\pm}$ are the phase shifts associated with the $L$ channels $L=J\pm1$ (when $L\neq J$) or with the spin channels $S=0,1$ (when $L=J > 0$).  Each $2\times 2$ block has a mixing parameter $\epsilon^{J\pi}$ that specifies the degree of channel mixing.  Since the $\delta_\pm$ are defined by eigenvalues of $\hat{S}$, they are independent of the angular momentum coupling scheme; $\epsilon^{J\pi}$ is not.  Converting the amplitudes of Eqs.~(\ref{Kaysmptote})-(\ref{Saysmptote}) from the $jj$ coupling of Eq.~(\ref{12wf}) to the customary $LSJ$ scheme for scattering is a straightforward exercise in Racah coefficients.

Because any pair of linearly independent solutions to Eq.~(\ref{coulombscrhodinger}) can be used to describe the asymptotic region in a given channel, it is straightforward to write the various scattering matrices in terms of each other \cite{newton1982book} as follows:

\begin{equation}
\label{StoT}
    \hat{T} = \frac{1}{2i}(\hat{S}-\hat{I})
\end{equation}
\begin{equation}
\label{StoK}
    \hat{K} = i(\hat{I}-\hat{S})(\hat{I}+\hat{S})^{-1},
\end{equation}
where $\hat{I}$ is the identity matrix.  We use all three of these. We apply the integral method to find the $A$ and $B$ amplitudes of the $K$-matrix formalism so that the work is formulated in terms of $F_l$ and $G_l$.  Our  phase shifts and mixing parameters are defined in the $S$-matrix formalism using Eq.~(\ref{blatt}).  Scattering experiments involve incoming plane waves, so that the $T$-matrix is the natural framework to compute differential cross sections.   We convert our initially $K$-matrix results to the other forms as needed.

\begin{acknowledgments}
We would like to thank R.\ B.\ Wiringa for being instrumental in the process of generating and applying the variational wave functions, M. Piarulli for the development and implementation of the Norfolk interactions, A.\ Kievsky for assistance in understanding scattering-matrix conventions, and M.\ Paris and G. Hale for explanations of the $R$-matrix phase shifts.  This work was supported by the US Department of Energy, Office of Nuclear Physics, Award DE-SC0019257.  
\end{acknowledgments}

\bibliography{vmcnT}

\end{document}